# On Synthesis of the Big Bang Model with Freundlich's Redshift and its Cosmological Consequences

## Asger G. Gasanalizade

*Shemakha Astrophysical Observatory,*

*National Academy of Sciences of the **Azerbaijan** Republic,*

*AZ 5616, Shemakha, P.O. Pirkuli, Azerbaijan*

E-mail: *gasanalizade@rambler.ru*

We derive exact theoretical value of the constant cosmic background radiation (CBR) temperature $T_0$ using the interconnections between the Gamow, Alpher and Herman (GAH) hot Big Bang cosmology model of the expanding Universe and the modified Freundlich redshift. As a result of this confluence an astonishing relationship between $T_0$ and the four fundamental physical constants $c$, $\hbar$, $k$, $G$ is found including also the Melvin's value of the Freundlich universal constant $A_S$. Then the resulting predicted the CBR temperature is $T_0 = 2.76625\ K$ .

This prediction show excellent agreement with the data obtained from ground-based and balloon-borne observations and also with a mean of the perfect black-body spectrum CMB temperature $2.725 K$ measured COBE in 1992. Using a new cosmological model we determine the horizon scale, age and mass of the present observable Universe. The calculations based on discrete redshift equations for the electromagnetic, electroweak phases and Planck epoch of the Universe predicts a graviton and string masses, which are originated beyond on Planck time. The predicted graviton mass $m_{Gr}$ is about five orders of magnitude less than the present "*the best possible upper bounds on the mass of the graviton*", which may be "*discovered*" in the proposed LISA observations. We present quantitative new results for the different quantum-cosmological parameters. Finally, it is showed that the mystery largeness and smallness dimensionless combination of the Quantum Cosmological constant $\Lambda_0$ and Planck length $l_{Pl}$ may be derived as their ratio from *the Trans-Plank redshift* relation. Thus is found the meaning a famous largeness cosmological number $c^3 / \hbar G \Lambda_0 = 2.8 \times 10^{125}$ that is inverse of $\Lambda_0 l_{Pl}^2 = 3.6 \times 10^{-126}$ , and "*which in 1930s was a regarded as a major problem by Eddington and Dirac*".

## Contents

## 1. Introduction

The Hot Big Bang model has become a principal part of the Standard Cosmological Model (SCM). However, one fundamental question such as what is the exact thermal cosmic microwave background (CMB), afterglow of the Big Bang as a very isotropic black-body radiation at a temperature of about $3K$, remains in this line of work for theorists (Turner, 1993; Bennet et al. 1996; Burbidge and Hoyle 1998; Peebles and Ratra 2003).The first modern treatment of the CMB radiation in the universe has been inferred (a small accuracy) by George Gamow and his co-workers Alpher and Herman (hereafter GAH) in the 1950s (Gamow 1948a,b, 1956; Alpher et al. 1948; Alpher and Herman 1948, 1950, 2001; see also, Omnes 1971, and references therein). But GAH model of the CMB radiation, cannot predict its correct absolute temperature. For example, they have determined the CMB temperature in different cases in the range from 5 to 40 $K$. Soon after, Finley- Freundlich (1954b) also called to fact that at the cosmological interpretation, his redshift relation predict two reasonable values of temperature for the cosmic radiation field given $T_S = 1.9K$ and $T_S = 6.0K$. Then these two comparable coincidence results implying that between the GAH theory and the Freundlich redshift hypothesis is exist the some a common and latent undiscovered logical connections. The succeeding cosmological search however, cannot to reveal the true meaning and correct magnitude of this duality. Thus, in spite of many years of studies (see e.g., Pecker 1991), there have not yet found any convincing explanation. As well known, SCM also failed to predict the exact *CMB temperature*. This was experimentally discovered accidentally a much latter by Arno Penzias and Robert Wilson (1965), which has come as a surprise to cosmologists. Their *"measurements of excess antenna temperature at 4080 Megacycles per second* "rise to a value that $T_0 = 3.1(1.0)K$. After first attempt a possible explanation, which has been made by Robert Dicke and his associates (1965), this fundamental discovery was rapidly accepted by the physical and astronomical community (see e.g., Weinberg 1977; Jastrow 1978 and references therein). This initial success of the experiment in the next few years stimulated numerous CMB temperature searches. Topics of the early history of CMB temperature measurements, followed by others may be found in Peebles (1971). In particular, from the statistical analysis of 17 independent microwave measurements completed in several years Peebles finds that thermo dynamical temperature is equal to $T_0 = 2.76(7)K$. From the weighted average of 18 published measurements Smoot et al. (1987) have derived $T_0 = 2.743(17)K$. This result is agree well with previous measurements reported by Weiss (1980) and Smoot et al. (1988) ( see also, Gush et al.1990; Halpern et al. 1991). More recently, a temperature $T_0 = 2.729^{+0.023}_{-0.031}K$ determined from a survey of CN rotational excitation temperatures in interstellar clouds (Roth et al. 1993).

The Cosmic Background Explorer (COBE) satellite observations (Smoot et al. 1992; Bennett et al. 1996; Fixsen et al. 1996; Mather et al. 1990, 1999) have shown that the CMB radiation is Planckian, at least, for the radio and microwave frequencies. Their corresponding values of CMB temperature are $T_0 = 2.728(2)K$ and $T_0 = 2.725(2)K$ at 95% confidence level (CL), respectively. Those measurements do not imply that the CMB spectrum is in form of the black body one at all redshifts, nor give information about how the temperature varies with redshift. However, in the recent is forced to accept the complementary compromise value of $T_0 = 2.725(1)$ K (Fixsen and Mather 2002).There is quite a large number of other experimental results on CMB temperature (Singal et al., 2006; Gervasi et al., 2008 and references therein) and this anisotropies (see e.g., Balbi et al., 2000; de Bernardis et al., 2000; Bennett et al., 2003 and references therein). The review and extensive list of bibliographic references relevant to the CMB are given in (White and Cohn 2002; Hu 2002).Need point out, however that, to this day were not available any data relating to the correct theoretical prediction of the CMB temperature. Therefore, it should be clear that, the problem remains to be solved. In a recent review Peebles and Ratra (2003) indicate that "the 3-K thermal cosmic background radiation is a centrepiece of modern cosmology, but its existence does not test general relativity".

We show below (Sec.4) how the unexpected interconnection between the Big Bang cosmology model and the Freundlich redshift hypothesis leads to the exact constant value of the cosmic background radiation (CBR) temperature, being in the excellent agreement with the observation of the Cosmic Microwave Background (CMB) temperature. The exact coincide of these temperatures opened the door to development of a comprehensive scheme to justify new redshift mechanisms with a quantum-relativity origin. By given nonflat model, for example, it is possible to make predictions the size, age and mass of the observable Universe. Most of the emphasis of this work will be on the cosmological redshifts that it goes back to the scales at just over Planck epoch. A cosmologically most relevant consequence of the epoch before the Planck epoch is that approaches the very earliest moments following the Big Bang there at redshifts $z >> 10^{62}$ and at temperatures $\sim 10^{94} K$ (Over Trans-Planck epoch) and $\sim 10^{63} K$ (Trans-Planck epoch) take place the processes at which are created the cosmological particles string and graviton, respectively (Sec.5 and 6). Especially important a role in our understanding of the early Universe is played the quantum value of cosmological constant which is well tested currently by experimental CMB temperature measurements (see, Sec.7). This model would have also a very clear picture of the explanation some underlying questions on the origin of the dark energy and the dark matter properties in the very early Universe. We will discuss some these possible problems and his consequences.

Below, we briefly review the GAH ideas on Big Bang cosmology model of the Universe and some of its consequences. More detailed discussions previous investigation can be found in the papers (Gamow 1948a, b, 1956; Alpher et al. 1948; Alpher and Herman 1948, 1950; ter Haar 1950; Omnes 1971).The consider revising of discovery and measurement of the CMB radiation in great detail presented also in Peebles (1971), Wilkinson and Peebles (1990) and Alpher and Herman (2001). For recent considerations of this subject see review (Peebles and Ratra 2003).

The motivation of Gamow's theory is based upon Tolman's ideas (1934) about the behaviour of radiation in the expanding Universe. Alpher et al. (1948) have applied Tolman's results to development of their own theory on the origin of the chemical elements in the early era of the Universe. However, at that time very little was known about the evolution of the Universe and the origin the lightweight chemical elements (Layzer 1984, Ch.8), so these calculated estimates were very uncertain, One important consideration of this idea allowed finding the approximate value of the CMB temperature, which as already mentioned, was subsequently discovered experimentally by Penzias and Wilson (1965).

## 2. Freundlich's redshift hypothesis

### 2.1 *Application to the solar spectrum*

In 1954, Finley-Freundlich (1954a, b) predicted that as a consequence of his "photon-photon interaction mechanisms" the wavelength $\lambda$ of every of absorption in certain types of B-, O-class stars (and solar) spectrum is displaced by $\Delta\lambda_F$ to the red with respect to the corresponding lines produced in a laboratory light-spectrum being defined by the equation

$$z = (\lambda_{obs} - \lambda_0)/\lambda_0 = A_F lT^4, \qquad (2.1)$$

where $\Delta\lambda_F = \lambda_{obs} - \lambda_0$, $\lambda_0$ and $\lambda_{obs}$ are the wavelength of a photon at emission and reception $(\lambda_{obs} > \lambda_0)$, $A_F = 2\times10^{-27} m^{-1} K^{-4}$ is the constant evaluated empirically from data on B – type stars, whose physical nature was obscure, $l$ is the path length traversed by a photon along which radiation density is constant and $T$ is the radiation field temperature. The solar spectrum redshift parameter $z$ is varying in the range $3\times10^{-7} < z < 4\times10^{-6}$.

Max Born (1954) was the first to realize the modern quantum electrodynamics meaning of the relation (2.1). He argued that the Freundlich redshift is a sequence of $N$ photon-photon collisions with a cross section which is equal to the square of the Compton wavelength of electron that is of about $10^{-25} m^2$. The last limit in the solar visual spectral range will be by the factor of $10^3$ smaller than expected (ter Haar, 1954a) and does not give $z \sim 3\cdot10^{-7}$ or $4\cdot10^{-6}$, but for the photon density of $\sim 10^{18} m^{-3}$ produces only a small change in the wavelength or frequency. Thus, the Freundlich redshift is linked to the radio astronomy. Born has shown that this hypothesis would indicate also the appearance of an absolute length in the field equations for vacuum, and as the laws of general relativity do not contain such a length they ought to be modified.

In the early times there was an extensive immediate discussion of the proposed hypothesis (Burbidge and Burbidge 1954; Ginzburg 1954; ter Haar 1954a, b; Helfer 1954; McCrea 1954; Struve 1954). Soon after, a more careful Melvin's (1955) calculations showed that the constant $A_F$ should be diminished by the factor of $10^3$. After this reduced estimation of the empirical constant $A_F$ the most criticisms of "Freundlich's mechanism" disappeared. In this connection it should also be mentioned that Freundlich himself did not impart a great weight to the formula (2.1), but put it forward only as a possibility to fit the various observational material which has not been explained by current astrophysical theories (ter Haar, 1954b). McCrea argue that "*the actual phenomena to which Freundlich directs attention call for much further investigations, both observational and theoretical*" (McCrea 1954; ter Haar 1954b).

Discussions in succeeding years (Neugebauer 1955; Browne 1962; Jorand 1962; Melnikov 1964; Melnikov and Popov 1974; Blum and Weiss 1967; Gasanalizade 1967, 1968, 1970, 1971; Pecker et al. 1972, 1973) also showed a continuous interest in this problem. Later, after of unambiguously identifications of the puzzling empirical parameter, so-called "interaction displacement" $Y$ (extracted by Freundlich and Forbes (1956a, b; 1959) using the least-squares fit of Adam's (1948) solar redshift measurements) with *a Einstein's of the gravitational redshift at the Lyman limit of hydrogen in solar spectrum* $\lambda_{Ls} \cong 1,93 \times 10^{-13} m$ (Gasanalizade 1967, 1968, 1970) and also coincidences of new cross section for photon-photon interaction in solar atmosphere spectrum $< \sigma_s > \cong 1,6 \times 10^{-24} m^2$ (Gasanalizade 1971) with the ter Haar's (1954a) predicted value of $\sigma_{pp} (\sim 10^{-24} m^2)$ criticisms of Freundlich's hypothesis not only stopped but this also assume over greater importance (Adam et al. 1976; Beckers and Nelson 1978; Beckers and Cram 1979; Dravins et al. 1981; Balthazar 1984; Pecker 1991; Gasanalizade 1992a, 1994). Thus, it is possible to say that the Freundlich mechanism at present form is acceptable hypothesis which well justified at least in a solar spectrum.

By comparing the GAH predictions with the Freundlich-Born hypothesis, Pecker (1991) concluded that*: "... It was indeed predicted with a weak accuracy, but nevertheless predicted, by Gamow, by Herman and Alpher, some then years before being observed: it was even looked for by Dicke, Peebles, Roll and Wilkinson, on the basis of the prediction, but not found, when found, more or less by change by Penzias and Wilson, Dicke recognized it as the radiation he was looking for, and the stage was set for a Nobel Prize. This indeed was a strange course of events! As everyone, in 1964, for got too easily that Finley-*

*Freundlich predicted in 1954, in a way better indeed than Gamow, so-called observation, as a consequence of the tired-light mechanism in a static Universe, and Max Born went along. Why have these papers been for gotten at the time of Penzias and Wilson's discovery? Hard to say! Possible of because Dicke was a convinced Gamowian? Perhaps only because Finley-Freundlich's and Born's papers were published in the Philosophical Transactions and the Proceedings of the Philosophical Society, much less distributed than Ap. J., Phys. Rev. or J. of Modern Physics?*

*My comment does not mean necessarily that we should agree with that tired light hypothesis, and reject the relic radiation interpretation; I just claim that things should be more open than they are".*

The history and drama of this discovery are described also in the references (Melchiorri and Melchiorri 1994; Partridge 1995, see also, White and Cohn 2002)

## 3. The temperature- time relation in the quantum cosmology

### 3.1. The meaning of quantum relativity cosmological constant $\partial$

According to the general relativity (see e.g., Alpher et al. 1948; Gamow 1956; Omnes 1971; Peebles 1971; Weinberg 1972; Silk 1980) a temperature-time variation of homogeneous isotropic hot Big Bang model of the expanding Universe is describe by the relationship

$$T \cdot t^{1/2} = (3c^2 / 32\pi aG)^{1/4} = 1.5 \times 10^{10} K s^{1/2}. \tag{3.1}$$

Here $c$ is speed of light in vacuum; $G$ is Newton's gravitational constant, $a = 4\sigma/c$ is the radiation density constant and $\sigma$ is Stefan-Boltzmann's constant. For the sake of simplicity, instead of the constant term $T^2 t$ we shall now denote it throughout by $\partial$ (first Azerbaijan small letter $\partial$ known as Latin letter "schwa"). In this context, we shall adopt in the rest of the discussion a slightly improved version as (Davies 1982, Ch.4):

$$\partial \equiv T^2 t = (45\hbar^3 c^5 / 32\pi^3 k^4 Gf)^{1/2}. \tag{3.2}$$

Here $\hbar$ is reduced Planck's constant $(\hbar = h/2\pi$, where $h$ is Planck's constant), $k$ is Boltzmann's constant and $f$ is the weighted dimensionless factor. Davies pointed out that value of $f$ is order of the unity. Indeed, if we set $f = 1$, we have the result conjectured in (3.1). The expression (3.2) may be simplified giving

$$\partial \equiv T^2 t = (c/k)^2 (3\hbar/4\pi)(5\hbar c/2\pi Gf)^{1/2}. \tag{3.3}$$

But the correct numerical value of constant $\partial$ from a given relation is derivable only by rather involved factor $f$. We find (Gasanalizade 2004) that the weighted dimensionless factor $f$ is given by

$$f = (5/2\pi)(\frac{3}{4\pi} \cdot 4.965114231)^2 = 1.11807620956.$$ (3.4)

Using this value for $f$, and at the numerical values $c = 2.99792458 \times 10^8 \, ms^{-1}$ (exact), $\hbar = 1.054571596(82) \times 10^{-34} \, Js$, $k = 1.3806503(24) \times 10^{-23} \, JK^{-1}$ (Mohr and Taylor 2000) and $G = 6.674215(92) \times 10^{-11} \, m^3 kg^{-1} s^{-2}$ (Gundlach and Merkowitz 2000), by virtue of Eq. (3.3) we obtain (Gasanalizade 2004, 2007)

$$\partial^{1/2} \equiv T \cdot t^{1/2} = 1.476334(23) \times 10^{10} \, Ks^{1/2}.$$ (3.5)

Then, the new quantum in nature and the trustworthy universal constant $\partial$ is equal to

$$\partial \equiv T^2 t = 2.179562(34) \times 10^{20} \, K^2 s.$$ (3.6)

(Throughout this paper the numbers in parentheses after the values give the one standard – deviation uncertainties in the last digits).

Eq. (3.2) and (3.3) indicate that the universal constant $\partial$ is similar to the Stefan–Boltzmann's constant $\sigma$. We find that Eq. (3.3) for $\partial$ may be rearranged to a physically equivalent and classical elegant form. Then, more precisely constant $\partial$ can be expressed as

$$\partial \equiv T_0^2 t_0 \equiv T^2 t \equiv \cdots = b(c_2 / c_1) E_{Pl}, \text{ where } c_1 / c_2 = 2\pi \cdot c \cdot k,$$ (3.7)

and where $b$ is Wien's displacement law constant, $c_1$ and $c_2$ are the first and second radiation constants, respectively and $E_{Pl} = (\hbar c^5 / G)^{1/2}$ is the Planck energy. Then, precise value of $\partial$ is free from the Davies's dimensionless factor $f$. Assuming the values of the Fundamental Physical Constants (FPC) from a CODATA-2006 data (Mohr et al. 2008) the new values of constant $\partial$ is given by

$$\partial = 2.17955_3(11) \times 10^{20} \, K^2 s.$$ (3.8)

As will be indicated in Sec.4 constant $\partial$ with the Freundlich–Melvin's constant (Freundlich 1954a, b; Melvin 1955) $A_S = 2 \times 10^{-30}\,\mathrm{m}^{-1}\,\mathrm{K}^{-4}$ can be used in many cosmological calculations, while $A_S$ is the constant depending on details of the some deeper laws that we have to understand. In Sec.5.3 is shown the application of this constant in particle physics reaches.

(In cosmological calculations the FPC usually expressed in terms of the normalized parameters at provided that $c = 1$ or in quite $c = G = \hbar = \kappa = 1$ etc., which for the physicist every so often can cause difficulties. Then, in this paper, and what follows, held the usual meaning of the each FPC).

As will be applied later the value $\partial$ is a universal constant for all concurrent combinations $T$ and $t$ after the onset of the Big Bang should be same, i. e.,

$$\partial \equiv T_0^2 t_0 \equiv T_1^2 t_1 \equiv \cdots \equiv T_{Pl}^2 t_{Pl} \equiv \cdots, \tag{3.9}$$

where the subscripts zero and $Pl$ denotes the values today and in the Planck time, respectively and defined below. This means that the term $\partial$ is one truly universal cosmological constant for all times of the Universe up to some cut off time $t_{Crit} \sim 10^{-169}\,s$ (see, Subsec.5.3.4). For example, from this equation it is clear that when observed CMB temperature is equal to $T_0 = 2.725(1)\,K$ (Fixsen and Mather 2002) we can conclude that the relevant age of the Universe today correspond to $t_0 = 2.935(2) \times 10^{19}\,s$.

### 3.2. "Number of particles – redshift duality"

Motivated by Freundlich hypothesis, for which cosmological redshift relation is given by Eq. (13) (Gasanalizade 2007) (see also Eq. (4.9) in the next Sec. 4) we rewrite Eq. (3.9) and introduce the new discrete time variable dimensionless parameter $N$ represented as

$$N(t_1) \equiv (\frac{T_1}{T_0})^2 \equiv \frac{t_0}{t_1} \equiv \cdots, \quad \text{and} \quad N(t_{Pl}) \equiv (\frac{T_{Pl}}{T_0})^2 \equiv \frac{t_0}{t_{Pl}} \equiv \cdots . \tag{3.10}$$

Physically this means that, $N$ is the number of "particles" which stays a fixed constants in the commoving volume in the early Universe in the course expansion (From here, so that distinct from the known physical particles, we shall call a particles $m_{Str}$, $m_{Gr}$, $m_{Pl}$ as a "*cosmological particles*". We will show later that these *three basic particles mass* can be regarded as a constant and related to the mass of the Universe $M_U$ by sequences as $m_{Str} : m_{Gr} : m_{Pl} : M_U \equiv N_{Pl}^{-1} \sim 10^{-62}$). The lightest cosmological particle in this model is the string mass weighting in at $m_{Str} \sim 10^{-133} kg \sim 10^{-98} eV/c^2$. In the course of expansion of the Universe, parameter $N$ is generated discrete. Its magnitude can be expressed as $N(t) \sim N^{\pm n}(t_{Pl})$, where $n = 0, 1, 2, 3$, and $N(t_{Pl}) \equiv N_{Pl} \sim 10^{62}$ is the number of "Planck particles" in the moment of Planck time $t_{Pl} \sim 10^{-43} s$ (i.e., at the Planck epoch).

A first consequence of Eq. (3.10) is that one can predict the number of "Planck particles" $N_{Pl}$ without even considering the origin of this number (This prediction is considered in more detail in the Sec. 5.4). In a second step one may use Eq. (3.10) in order to estimate the number of other particles originating in different phases of expansion of the Universe. Particularly, in case particle physics at $t_0 > t >> t_{Pl}$, one can have $N(t_0) < N(t) << N(t_{Pl})$, where $N(t_0) \equiv N^0(t_{Pl}) \equiv N_0 \equiv 1$. Thus, Eq.(3.10) links the number of cosmological particles $N(t)$ to a cosmic time of this creation in early Universe by

$$t_0 \equiv \partial / T_0^2 \equiv N(t_1)t_1 \equiv \cdots \equiv N(t_{Pl})t_{Pl} \equiv \cdots . \qquad (3.11)$$

This phenomenon can be interpreted as the space-time particle-redshift duality, because below it identified with the modified Freundlich redshift parameters $z_{Fr}$. In what follows these numbers of any particles will be determined by the Freundlich redshift relations. Then, *the centrality of the point particle* will be *replaced by the centrality of the wave*-redshift (Gates 2006). Here this conformity for the moment will be omitted and we will return to this question in Sect. 4. However, it is important to note that we may consider of Eq. (3.10) as a solution of the existing puzzle mentioned in Introduction as the hidden duality. As will be shown below the essentially identity $N(t)$ and $z_{Fr}(t)$ by setting $N(t) \equiv z_{Fr}(t)$ reflects the coincident alternating relation between quantum relativity and the modified Freundlich redshift equation. In other words, we find a model in which is established duality of modified Freundlich redshift and numbers of cosmological particles. Thus, this equivalent identity is one of great triumphs of synthesis named in a title paper.

### 3.3. A brief survey on time variability of the FPC

In above equations, and in that what follows the effects of cosmological change of Newton's of gravitational constant $G$ (see e.g., Brans and Dicke 1961; Wesson 1973; Barrow 1978, 1987; Gasanalizade 1992b, 1994) and other FPC constants have been neglected (see also, Subsec.5.4). On the other hand, a combination of experimental observations and theoretical analyses in 2000's led to conclusion (Bambi et al. 2005; Will, 2006, and their references), that for time dependence of $G$ could a bound on $(\frac{dG}{dt} \cdot G^{-1}) < 10^{-13} yr^{-1}$. As pointed out by Uzan (2003) the cooling process in white dwarfs is accelerated if $(\frac{dG}{dt} \cdot G^{-1}) < 0$.

Assume a Big Bang nucleosynthesis (BBN) model for evolution, Copi et al. (2004) deduced that present limit on the time variation of $G$ today is sufficiently small and can be estimates as $-3 \times 10^{-13} \, yr^{-1} < (\frac{dG}{dt} \cdot G^{-1})_{today} < 4 \times 10^{-13} \, yr^{-1}$. Thus, we have verified that the constant $G$ over an above range $(\sim 16 \times 10^{-6})$ is consistent with no variation in this quantity. By Uzan (2003) the "*hypothesis of constancy of the constants plays an important role in particular in astronomy and cosmology where redshift measures the look-back time. Ignoring the possibility of varying constants could lead to a distorted view of our universe and if such a variation is established corrections would have to be applied. It is thus of importance to investigate this possibility especially as the measurements become more and more precise.*"

Idea on possible time depending variability of other physical constants in the expanding universe has already been discussed by Dyson (1972), Yahil (1975), Baum and Florentin-Nielsen (1976), Solheim et al. (1976), Bekenstein (1982), Marciano (1984), Kolb et al. (1986) and Cohen (1987). In particular, a much less stringent limit on relative change of fine structure constant $\alpha$ per year must be less than $10^{-16}$ per year (Bahcall and Schmidt 1967; Wolfe et al. 1976; Damour and Duson 1996; Webb et al. 1999; Avelino et al. 2001; Levshakov et al. 2002). Murphy et al. (2003) using very quality data for Si IV doublets found $\Delta\alpha / \alpha = (-0.57 \pm 0.10) \times 10^{-5}$ over the redshift range $0.2 \le z \le 3.7$. Bahcall et al. (2004) using OIII emission lines from QSOs have found $\Delta\alpha / \alpha = (0.7 \pm 1.4) \times 10^{-4}$. Chand et al. (2005) have shown that the variation of the fine-structure constant based on the analysis of 15 Si IV doublets is equal to $\Delta\alpha / \alpha = (+0.15 \pm 0.43) \times 10^{-5}$ over the redshift range $1.59 \le z \le 2.92$ which is consistent with no variation in $\alpha$ (see Uzan 2003; Will 2006, for a very complete review).

For explain the horizon problem the varying speed of light (VSL) hypothesis has been proposed by Moffat (1993), Albrecht and Magueijo (1999) and Barrow and Magueijo (1998; 1999) under the claim that, this offers an alternative way of solving the SCM problems. In a recent study on black holes it is suggested (Davies et al. 2002) that variation of the speed of light can be discriminated from a variation of the elementary charge. This problem has been considered also by Carlip (2003). Barrow (2003) proposed that at continuous increase of speed of light the time dependent speed of light varies as some power of the expansion scale factor $a$ in such way that $c(t) = c_0 a^n$, where $c_0 > 0$ and $n$ is constant. The time variation of $c$, $G$, and $h$ are considered also by (Uzan 2003; Buchalter 2004). More recently, the theoretical motivations for such variation as estimated by Ellis and Uzan (2005), and Will (2006) ( see also, Fritzsch 2009) are focused mainly on the possibility of VSL hypothesis, fine structure constant $\alpha$, Planck constant $h$ and Newton's constant $G$ time-variations.

Notice that, the validity of the $T$ and $t$ from relationship $\partial \equiv T^2 t$ and constant $A_S$ could be questionable only for very early times of Universe, i.e., in the vicinity of the singularity at $0 \le t << t_{Pl}$ where $t_{Pl} = (\hbar G/c^5)^{1/2} = 5.39 \times 10^{-44} s$ is the Planck time (Ginzburg 1971; Schramm 1983; Lieu and Hillman 2003; Gasanalizade 2004). Glashow (2002) called the time $t < t_{Pl}$ the "*taboo for a physicist*". Nevertheless, the law $\partial \equiv T^2 t$ admits that the lower limit of cosmic time $t$ in the very early Universe may be drastically reduced on the order $t_{Crit} \sim 10^{-169} s$ after the Big Bang (see Eq. (5.20c)).This is very much smaller than Planck time. However, Carr (1980) argues that Planck time, *"after all, is the beginning of the classical universe"*.

## 4. Modified Freundlich redshift in cosmology

We now turn to our main problem connected with renewed SCM by calculation of the Freundlich redshift once synthesis this with those. Based on our present knowledge, Freundlich's hypothesis in his original form is not acceptable in the current cosmology (Pecker et al. 1972, 1973; Woodward and Yourgrau 1973; Aldovandi et al. 1973) and bound to be modified in the light of the more sophisticated theories. We shall here use a modification of Freundlich's formulation in which cosmological redshift maybe classified as $z \ge 1$ (Carr 1980; Webb et al. 1999). As a result, this modification maybe considered as a revising metric of the space-time model, which satisfies Einstein's equations. This model mathematically can be reconstructed following way (most likely that in this case at frequencies lower than 10 $GHz$ the Fermi-Dirac gaseous model were replaced by Boze-Einstein gaseous model, see e.g., Gervasi et al. 2008)

$$z_{Fr} = A_S l T^4 = c A_S T^2 (T^2 t).$$ (4.1)

Here, $A_S = 2 \times 10^{-30} m^{-1} K^{-4}$ is the Melvin (1955) theoretical value of the Freundlich universal empirical constant, $l = ct$ is the distance travelled by a photon since the beginning of the expansion of the universe (i.e. event horizon), $t$ is the cosmic expansion time equal to the lifetime of the universe after the Big Bang, and $T$ is the CBR temperature. There the value $T^2 t$ in brackets is the constant, which does not determined by Freundlich's hypothesis. This is a uniquely defined constant $\partial$ that was directly calculated above (see Eqs. (3.2)–(3.9)) from the Big Bang cosmology model. Now, at given values of the constants $c, A_S, \partial$, redshift $z_{Fr}$ essentially proportional to the square of $T$ and inverse proportional to the parameters of $t$ and $l$. In addition, modified Freundlich redshift $z_{Fr}$ as an alternative to Eq. (3.10) takes the convenient form

$$z_{Fr} \equiv (\partial c A_S) T^2 \equiv (\partial^2 c A_S)/t \equiv (\partial c)^2 A_S /l \equiv (\partial^2 A_S) g \equiv \cdots .\qquad (4.2)$$

Here $g$ is a parameter that interpreted as a background cosmic acceleration. In a preceding work (Gasanalizade 2007) we have already mentioned that, when present cosmic redshift become as $z_0 = (z_{Fr})_0 = (c\partial A_S) T_0^2 = 1$, from the relation (4.2) can be reduced the formulas for the today epoch of the Universe parameters given by

$$T_0 \equiv (\partial c A_S)^{-1/2}, \ t_0 \equiv \partial^2 c A_S \ , \ l_0 \equiv (\partial c)^2 A_S \ ,\qquad (4.3)$$

where the subscript zero denotes that these predicted parameters are for the cosmic background temperature, the age and the event horizon values of the Universe at the present epoch, respectively. It should be noticed that the above relation (4.2) predicts also the constant acceleration by the Freundlich's cosmological redshift, given as

$$g_0 \equiv c/t_0 \equiv c^2/l_0 \equiv (\partial^2 A_S)^{-1},\qquad (4.4)$$

where $g_0 \sim 10^{-11} m/s^2$ defined as the extremely least constant limit for the cosmic ("relict") background acceleration for a today Universe. In "*classical gravity*" this parameter is known as "*acceleration due to gravity*" (the Newton expression for acceleration) and can be expressed as

$$g_0 \equiv g_{grav} \equiv G M_U /R_U^2 \ .\qquad (4.5)$$

It is interesting to note that, before Einstein (1917) named $M_U$ as "*characteristic mass-energy")*, and $l_0 \equiv R_U$ as the "*characteristic length scale*" (Will 2006) of the phenomenon given by

$$(G/c^2)(M_U /R_U) = 1.\qquad (4.6)$$

These parameters are presented before (Gasanalizade 2007) in the form

$$M_U \equiv (c^2/G)l_0 \quad \text{and} \quad R_U = l_0. \tag{4.7}$$

In this particular form, the current horizon size $l_0$ is not arbitrary, but in fact precisely equals the physical radius of the Einstein's Universe with characteristic mass-energy corresponding to a closed model. In addition, the quantity of the today background acceleration $g_0$ can be defined as

$$g_0 \equiv c/t_0 \equiv c^2/l_0 \equiv GM_U/R_U^2 \equiv GM_U/l_0^2 \equiv (\partial \cdot A_S)^{-1}. \tag{4.8}$$

Combining Eq. (4.2) with Eq. (4.3), we find (Gasanalizade, 2007)

$$z_{Fr} \equiv (\frac{T}{T_0})^2 \equiv \frac{t_0}{t} \equiv \frac{l_0}{l} \equiv \frac{a}{g_0} \equiv \cdots, \tag{4.9}$$

which is still fully compatible with relation (3.10) derived for the hot Big Bang model. However, in this model $T_0$, $t_0$, $l_0$ and $g_0$ are current quantum gravity origin constant parameters of the Universe. There $t_0$ characterizes the evolutionary today timescale of the universe after the Big Bang ("*the current age of the universe*") and $l_0$ is the largest length scales currently accessible in cosmological observations ("*the current horizon size*" $= ct_0$). The parameter $a$ in epoch $z_{Fr}(t)$ of the Universe is the cosmic acceleration. At the present epoch $z_{Fr} \equiv z_0 \equiv 1$, and $T \equiv T_0$ is the cosmic background radiation (CBR) temperature.

From Eq.(4.9) and (3.10) it can be read off immediately that the expressions in the right part of these relations are same, i.e., $(T/T_0)^2 \equiv t_0/t$. This implies that $N(t) \equiv z_{Fr}$ and then $N_0 \equiv z_0 \equiv 1$. Next we combine these data. Then the synthesis of modified Freundlich's redshift hypothesis and the Big Bang cosmological model acquires a consistent physical basis. Hence Eq. (4.9) has a quite general meaning on the redshift and time and may be applied to the different epochs of the Universe after the Big Bang. The simplest bound for the universe model is based on the assumption that for all the early epoch redshift $z(t) > z_0 \equiv 1$ classified as "*cosmological*" (Carr 1980).

In present paper, within a certain approximation, which is described in detail below, to probe the redshift-space time relationship in the early Universe expansion rate, we use a small number of the new basic quantum cosmological parameters $T_0$, $t_0$, $l_0$ and $g_0$ plus the additional parameters noted below in the text. These parameters uniquely determine the model of quantum processes in the beginning of the Universe, leads to a solution of the cosmological constant problems and have revealed the presence of graviton and string masses. Another interesting feature of this model is that the temperature dependence on the redshift $z$ of the relic radiation $T(z)$, is some different from the SCM type conception. As is seen, there the value for Freundlich's redshift $z_{Fr}$, determined directly from the (4.2) and (4.9), is proportional to the square of the CBR temperature.

An alternative explanation of the accelerating expansion rate of the Universe is the Friedmann–Lemaitre model, which is determined as first power function of CMB temperature by relation $1 + z = T / T_0$ (see e.g., Peebles 1971, Ch. 7). This is not throughout checked by direct substitution. In this paper we deduce a quadratic law $z_{Fr} = (T / T_0)^2$, what radically departure from of the conventional scenario, where redshift dependence of the CMB temperature is linear $T(z) = T_0(1 + z)$. In the case (4.9), redshift is most sensitive to the ratio $(T / T_0)^2$ and greatly ameliorated redshift dependence of the CMB temperature problems in the expanding Universe.

In footnote (p. 585) of a recent review Peebles and Ratra (2003) pointed out that "*we do not know the provenance of this (i.e., $T \propto (1 + z)$) argument; it was familiar in Dicke's group when the 3-$K$ cosmic microwave background radiation was discovered*". In other work Peebles (2000) write:"…*Neil Turok considers what the universe might have been likely at redshifts so high the Friedmann-Lemaitre model certainly could not have applied. Turok very correctly emphasizes that the question is open and absolutely must be addressed. But we have to live with the fact that an empirical validation of the answer may be a long time coming. …The empirical basis for research on the early universe is a lot more limited*". Then, it can be inferred that in this question present a dubious problem. Nevertheless, in present case, in which are known two formulation of the redshifts the Freundlich redshift would be obtained as the square of values $(1 + z_{F-L})$ by

$$z_{Fr} = (1 + z_{F-L})^2, \tag{4.10}$$

where $z_F(t)$ and $z_{F-L}(t)$ are the Freundlich and Friedmann-Lemaitre redshifts, respectively. Therefore, transpire that when redshift passing from the lowest values $z \leq 1$ to the higher values $z > 1$ this varies with the square of the ratio $(T / T_0)^2$.

There is no more closely observed experimental cosmological parameter than the CMB temperature (see e.g. Alpher and Herman 2001; Peebles and Ratra 2003; Will 2006, and his references). In recent year the value of the CMB temperature has been measured with increasing accuracy (see, Subsec.5.2.1). Since the CMB temperature is of such fundamental importance we believe that the predicted CBR temperature in a few parts in $10^{-3}$ is coincides one. Then, below the CBR temperature $T_0$ will be more specifically interpreted as $T_0(CMB)$, although this can be extended to an all wavelength (or frequency) as the thermodynamic temperature. This observable $T_0$ gives a more direct access to the Freundlich redshift for the following applications. In present work owing to the new conception, graviton mass $m_{Gr}$, today background acceleration of the Universe $g_0$, the cosmological constant $\Lambda_0$, the Universe mass $M_U$ and other cosmological parameters are expressed at once in terms of the recent experimental CMB temperature (Singal et.al 2006) and determined ( see  e.g., Eq.(5.3), (6.3) and (7.8) and table 3).

## 5. Application of the Freundlich redshift to an expanding universe model

### 5.1. The present epoch of the Universe

In table 1 we have summarized our results and give numerical values of cosmological parameters assuming that today redshift is $z_0 \equiv 1$, and value of $\partial$ defined by Eq. (3.8) and (3.9) at $cA_S = 5.99584916 \times 10^{-22}\, s^{-1} K^{-4}$. Evidently, within the SCM there is no way to calculate these parameters. Motivated by the aesthetics of a serious problem we include also cosmological constant $\Lambda_0$ with a dimensions of $[length]^{-2}$ (Abbot 1988). As in what follows we us $\Lambda_0 = l_0^{-2} = 1.37 \times 10^{-56}\, m^{-2}$.This result is a much smaller than the current value of the cosmological constant $\Lambda_0 \cong 10^{-52}\, m^{-2}$ (Peebles and Ratra 2003). (Cosmological constant is detailed in Sec. 7).

**Table 1**.The today quantum relativity cosmological parameters of the Universe

| Quantity, Symbol | Definition | Value |
|---|---|---|
| **Background temperature, $T_0$** | $(c \cdot \partial \cdot A_S)^{-1/2}$ | $2.76625\ K$ |
| Age of the Universe, $t_0$ | $c \cdot \partial^2 \cdot A_S$ | $2.848 \times 10^{19}\, s$ |
| The horizon size, $l_0$ | $(c \cdot \partial)^2 \cdot A_S$ | $8.539 \times 10^{27}\, m$ |
| Background acceleration $g_0$ | $(\partial^2 \cdot A_S)^{-1}$ | $1.0524 \times 10^{-11}\, m/s^2$ |
| Cosmological constant, $\Lambda_0$ | $= l_0^{-2}$ | $1.37 \times 10^{-56}\, m^{-2}$ |

### 5.2. The consistence with the observational results

#### 5.2.1. Cosmic background radiation temperature

Calculations by the relation (4.3) for the CBR temperature $T_0$ have showed that $T_0 = (c\, \partial\, A_S)^{-1/2} = 2.76625K$ , which is in an excellent agreement with all of the available data obtained from ground-based and balloon-borne observations $(2.76 - 2.90K)$ (Weiss 1980; Smoot et al. 1987; Sironi and Bonnelli 1986; Johanson and Wilkinson 1987; Smoot et al. 1988; Matsumoto et al. 1988; Kaiser and Wright 1990; Meyer et al. 1989; Crane et al. 1989; Palozzi et al. 1990; Gush et al. 1990; Halpern et al. 1991; Battistelli et al. 2002).

The predicted value for $T_0$ exceeds the latest COBE measurements of the CMB temperature $(2.725(2) - 2.728(2)K)$ (Bennett et al. 1996; Fixsen et al. 1996; Mather et al. 1999) by 56 $u_{diff}$, where $u_{diff}$ is the standard uncertainty of their difference and hence the two values are in the severe disagreements. The basic results derived from the first-year observations of the Wilkinson Microwave Anisotropy Probe (WMAP) team (in Bennett et al. 2003, and references therein) have supported a largely distributed black body radiation temperature as $2.725\,K$ with random anisotropies of only a few parts in $10^{-5}$ over the whole sky. (For a review of bibliographic references to the CMB see also, White and Cohn 2002).

Predicted CBR temperature also fully coincides with the recently derived CMB temperature which has been measured in frequency range from 3 to 90 $GHz$ by the Absolute Radiometer for Cosmology, Astrophysics and Diffuse Emission (ARCADE 2) experiment (Singal et al. 2006).These experimental CMB temperature are given as

$$T_0(\text{CMB}) = (2.766 \pm 0.160)K \quad \text{at} \quad 8.3\,GHz, \tag{5.1a}$$

$$= (2.897 \pm 0.116)K \quad \text{at} \quad 8.0\,GHz. \tag{5.1b}$$

In another recent work, Gervasi et al. (2008) consider the results of measurements the absolute temperature of the CMB in the lower frequency region obtained

$$T_0(CMB) = (2.738 \pm 0.129 \pm 0.066)K \quad \text{at} \quad \nu = 0.60 \text{ GHz}, \tag{5.2a}$$

$$= (2.803 \pm 0.051^{+0.430}_{-0.300})K \quad \text{at} \quad \nu = 0.82 \text{ GHz}, \tag{5.2b}$$

$$= (2.516 \pm 0.139 \pm 0.284)K \quad \text{at} \quad \nu = 2.5 \text{ GHz}. \tag{5.2c}$$

Let us note that the main proposal made here is that the theoretical predicted CBR temperature $T_0(pred)$ is fairly close coincident with the values derived from the last mentioned experimental measurements CMB temperature. Further, this makes a strong prompted that, the synthesis of the Freundlich redshift and Big Bang cosmological model, as whole, is amenable to the requirements of the Bose-Einstein quantum statistics.

Finally, it should be noted that it is the rarely occurred case in astrophysics and cosmology. Since theoretical value of $T_0(pred)$ depends exclusively on the FPC and also the Melvin's value of the Freundlich universal constant, this result has opened a new research field in the cosmological metrology. The problem, however, is that the Freundlich-Melvin's value constant $A_S$ nevertheless provides a reasonable approximation for the onset of determination cosmological parameters $T_0$, $t_0$, $l_0$ and others, but the precision of one produce of a special concern. Then, although in the $T_0(\exp)$ measurements may well be various other undetectable distortions ( Parijskij 1973; Kogut et al. 1988; Gush et al. 1990; Turner 1993; Bennett et al. 1996; Fixsen et al. 1996; Mather et al. 1999; Fixsen and Mather 2002; Fixsen et al. 2004), we use recent ARCADE data (Singal et al. 2006) to derive a somewhat stronger limit for adjusted value of constant $A_S$, assuming $T_0(CMB)$ given in of (5.1a) and adopting their error analysis. By taking

$$T_0(CBR) = (\partial c A_S)^{-1/2} = T_0(CMB) = 2.766 \pm 0.160 K \,, \tag{5.3}$$

we have a new refined value of $A_S$ which can be written in the form

$$A_{S0} = (2.000 \pm 0.116) \times 10^{-30} m^{-1} K^{-4} = 2.00(12) x 10^{-30} m^{-1} K^{-4} \,. \tag{5.4}$$

Then, the numerical uncertainties in calculations of cosmological parameters are $\sim 5.8 \times 10^{-2}$, while future measurements will lead to a reductions just as in the uncertainty of $A_{S0}$, so also other parameters. Because of such correction of the constant $A_{S0}$ values of $t_0, l_0, g_0, \Lambda_0$ and other predicted parameters can be probed within the above accuracy. Then

$$t_0 = (2.85 \pm 0.16) \times 10^{19} s = 2.85(16) \times 10^{19} s \,, \tag{5.5a}$$

$$l_0 = (8.54 \pm 0.49) \times 10^{27} m = 8.54(49) \times 10^{27} m \,, \tag{5.5b}$$

$$g_0 = (1.052_4 \pm 0.61) \times 10^{-11} m/s^2 = 1.052_4(61) \times 10^{-11} m/s^2 \,, \tag{5.5c}$$

$$\Lambda_0 = (1.37_1 \pm 0.16) \times 10^{-56} m^{-2} = 1.37_1(16) \times 10^{-56} m^{-2} \,. \tag{5.5d}$$

This estimate of $A_{S0}$ offer considerable scope for further improvements to obtain an accuracy of cosmological parameters that is used for all other calculations discussed below.

*5.2.2. The age of the Universe*

It has long been recognized that if age of the Universe can be determined by the flat cosmological model with $\Lambda = 0$, then this meant that Universe is younger than it oldest stars. Sandage's (1961) analysis indicates that this problem is removed by adding a $\Lambda > 0$. The result derived from the WMAP first-year observations under the flat space-time assumptions defined an age of the Universe as $t_0 = 13.7(2)\,Gyr$ (Bennett et al. 2003). However, in the context of the Eq. (5.5a) the predicted age of the current Universe corresponds to an $9.03(52) \times 10^2\,Gyr$. This is about a 70 times of magnitude older than above recent estimates as well as for an Einstein-de Sitter cosmology model age ($\Lambda = 0$ and $t_0 = (2/3)H_0^{-1}$) $13\,Gyr$. Nevertheless, the other cosmological models predicted that the Hubble age of the Universe $H^{-1} < 25\,Gyr$ (Sandage and Tamman 1990), whereas the globular clusters, which are object of the Universe exhibits the age of least from 15 to 19 billion years old (Iben 1974; Sandage 1982; Jakson 1992; Pont et al. 1998) or even older. "*The dynamical age*" $t_0$ (the inverse of the Hubble constant ) of the present-day Universe by Riess et al.(1998) is equal $14.2^{+1.0}_{-0.8}\,Gyr$ that is very close to those by Perlmutter et al. (1999) result $t_0 = 14.9^{+1.4}_{-1.1}$ (0.63 h) $Gyr$ for a flat space-time. That at present by the SCM the age of the some stars larger than the predicted age of the Universe. Thus, it is obvious that our prediction is removed the current conflict over the age of the Universe (Bolte and Hogan 1995; Pont et al. 1998; Lopez-Corredoire 2008). In particular, at the present epoch if the Hubble constant is assumed to be $H_0 = h(9.77813 \times 10^9\,yr)^{-1}$, where $h = 0.71(7)$ (PDG 2002), the our value of dimensionless age parameter (the Hubble time–units) $H_0 t_0 = 1.24$ in the $3\sigma$ level coincides with the upper limit of $H_0 t_0^{flat} = 0.96^{+0.09}_{-0.07}$ from the 42 high-redshift supernovae data of Perlmutter et al. (1999). However, our analysis was performed under the assumption that the present value of the cosmological constant is very small, but not zero (see Table 1). Another WMAP team surveys (Spergel et al. 2003; 2007) also suggest $\Lambda_0 > 0$. Thus, precision cosmological experiments indicate that the cosmological constant is nonzero, positively value (Sahni and Starobinsky 2000) with the magnitude $\Lambda_0(G\hbar/c^3) \approx 10^{-123}$ (Padmanabhan 2003). Peebles and Ratra (2003) pointed out that *"the data require $\Lambda_0 > 0$ at two to three standard deviations depending on the choice of data and method of analysis"*. (More on results about the $\Lambda_0(G\hbar/c^3)$ will be discussed in Sec. 8).

*5.2.3. On the topology of the Universe*

The first year background data collected by the NASA satellite WMAP (Bennett et al.2003; Spergel et al. 2003) has recently produced a high resolution, low noise maps of the temperature fluctuations in the CMB radiation. Then, in a nearly flat spherical case, with a density parameter $\Omega_0 = 1.02 \pm 0.02$ at the $1\sigma$ can be provide a powerful tool to derivation the other cosmological parameters of the Universe. In addition, recently, on the WMAP data Cornish et al. (2004) search of nearly-antipodal matched circles on angular sizes greater than $25^\circ$ the sky with similar temperature patterns to place constraints of the topology of the universe. For a spatially flat or nearly flat ($\Omega \approx 1$) model is found that our observable Universe have the diameter smaller than 24 $Gpc$ (1 $pc = 3.0857 \times 10^{16}$ $m$). It is expected that from WMAP's extended mission, or new data from the European satellite "Planck Surveyor", which are purposed to produce much higher resolution and much more sensitive CMB temperature fluctuation, measurements can be able to search for smaller circles and extended the limit to ~ 28 $Gpc$. By Cornish et al. (2004) if the Universe larger than this limit, then the circle statistic will not be able to constrain it shape. In connection with this it is interesting to compare these "observable" limits with the theoretically predicted value of today Cosmological event horizon of the Universe (roughly the present radius of the entire observable Universe) $l_0 = 8.54(49) \times 10^{27}$ $m = 5.28 \times 10^{62} l_{Pl}$, where $l_{Pl} = 1.616252(81) \times 10^{-35}$ $m$ (Mohr et al. 2008) is the Planck length. In SI units 24$Gpc$= $7.41 \times 10^{26}$ $m$ and 28$Gpc$= $8.64 \times 10^{26}$ $m$. Then, the predicted size of the Universe based on the above new redshift law exceeds the assumed from the WMAP data "observable diameter" of the today Universe by a factor $l_0 \approx 10 \times 28 Gpc \approx 277 Gpc$. Hence $d_{obs} \approx 28 Gpc$ to be lower than predicted value by more than on order of magnitude. As reviewed above the predicted $l_0$ is the distance that light has been able to travel during the time $t_0 = l_0 / c = 2.85(16) x 10^{19}$ $s \approx 9.03 \times 10^{11}$ $yr$ since the Big Bang expansion began.

Luminet (2005) calculated the space curvature radius by the Poincare Dodecahedral Space model (Luminet et al., 2003) also on the basis of WMAP data. In suggested model the curvature is (slightly) spherical, rather than by an effectively infinite flat model and when the WMAP data are better matched by a geometry. For the total energy density $\Omega_0 = 1.016$ (the WMAP data give: $1.02 \pm 0.02$), of the dark matter density $\Omega_m = 0.28$, of the Hubble constant $H_0 = 62 km / s / Mpc$, and the radius of the last scattering surface $R_{lss} \sim 53 Gpc$ the radius of space $R_c$ is $R_c = 2.63 R_{lss}$. Together with the value of 1 $Gpc$ this implies

$$R_c \approx 139.4 Gpc = 4.3 \times 10^{27} m. \tag{5.6a}$$

Thus, for the ratio $l_0 / R_c$ we obtain

$$l_0 / R_c = 1.985 \approx 2. \tag{5.6b}$$

Then, in contrast to analysis by Cornish et al. (2004), result suggested by Luminet (2005) for the radius of the observable Universe is twice as smaller than the predicted value.

There exists an alternative view of the Universe radius, just discussed by Adler et al. (2006). From truncate the matter at a commoving radius to form a finite limit on the mass and size of Universe model these authors deduced that at the present Hubble distance of about 14 $Gly$ ($1Gly$ is equal to $9.461 \times 10^{24} m$) the most bound for the Schwarzschild radius $R_u$ corresponds to

$$9.7 R_d \text{ (Lowest bound)} \leq R_u \leq 39 R_d \text{ (95 \% C L.)}, \tag{5.7a}$$

where $R_d$ is the de Sitter radius, which is related to the cosmological constant by $R_d = (3/\Lambda)^{1/2}$. Since by (5.7a) $R_d \approx 16 Gly$ we find $160 Gly \leq R_u \leq 620 Gly$. Thus, assuming $< R_u > = (160 + 620) Gly / 2 = 390 Gly = 3.69 \times 10^{27} m$, the ratio between our value of $l_0$ and the mean values for the Schwarzschild radius (The lower limit determined by Adler et al. (2006) is about $50 Gly$) should of

$$l_0 / < R_u > \cong 2.3. \tag{5.7b}$$

Then, the two determinations of the radius of the observable universe are roughly comparable. Actually, the recent observations of CMB measurements have established that *the large-angular scale (Sachs-Wolfe effect) plateau (l<100) in the angular power spectrum arises from perturbations with periods longer than the age of the universe at last scattering, i.e. ~ larger than the horizon, scales that can be affected by causal physics at that time* (White and Cohn 2002). Then, the "*larger than the horizon scale length*" most closely corresponds to the edge of our observable Universe with size of $R \sim 10^{27} m$ (Tegmark 2003). Eventually the relevant size of a flat plateau corresponds to the length mentioned above cosmological parameter $l_0$. This also means that local cosmological constant given as $\Lambda_0 = l_0^{-2}$ would be detectable in the present era (see also, Bousso 2006).

### 5.3. Previously epochs of expanding Universe

We now turn to a new dominant point, a radical change in what we accept as a persuasive theoretical foundation for a cosmological redshift in a very early epoch of the Universe. Unfortunately, in the SCM the situation with massive particles and energy dependence on cosmic redshift is not clearly known. This point is discussed by V. Icke (2004). His critical article concludes as: "*Just in case that you think this is a trivial game, pleas remember that there is currently no theory that describes the cosmic redshift in terms of an actual coupling mechanism between particles and space-time! And it would definitely have observational* *consequences; for example, the gravitational lens effects might not be achromatic. You would think that the quantum mechanical description of the cosmic redshift, so very well observed, would be the first thing that people would try to crack. But how? If I knew, I would be partying in Stockholm next year. Or you would, if you knew a better way."*

Then these are good reasons for considering of creation mechanisms some physical particles and acquire it a mass after the Big bang as a function of cosmological redshift. In realistic scenarios this period of early Universe corresponds to the end of inflation (Giudice et al. 2005). In principle, the correct moment of copiously creation of several particles and the gain of masses, followed from the (4.9).

*5.3.1. The quantum relativity redshift between particles and space-time*

We infer from above content that, starting particles most responsible for the quantum cosmology describe of redshift may be an electron, $Z^0$ boson and it is possible graviton, with the spins 1/2, 1 and 2, respectively, and which are to obeys the rule of the different statistics.

   *5.3.1.1. Redshift in the electromagnetic phase*

In the present and next Sections we focus our attention to the behaviour of physical parameters of the electron and $Z^0$ boson as a function of the space-time and high redshift in the electromagnetic and electroweak phases, assuming $(t_e \sim 10^{-21} s)$ for electron and $(t_{Z^0} \sim 10^{-26} s)$ for $Z^0$ boson, respectively. For definiteness, we express the value $t$ in terms of a unit of time fixed by the fundamental constants of atomic physics, say the electromagnetic time scale in natural units (n.u.) (Mohr et al. 2008) $\hbar / m_e c^2$, where $m_e$ is the electron mass. This time in laboratory scales is defined as the duration which is necessary for light to cover a distance equal to the Compton wavelength $\lambdabar_e = \hbar / m_e c = 3.8615926459(53) \times 10^{-13} m$ (Mohr et al. 2008) of an electron. By analogy of Dirac's (1937, 1938) principle of large numbers (LN), this implies that at the electromagnetic time phase of $t_e = \hbar / m_e c^2 = 1.2880886570(18) \times 10^{-21} s$ in an isotropic Universe a redshift $z_e \equiv N(t_e) \equiv N_e$, according to (4.9) can be expressed through fundamental constants of particle physics and quantum cosmological parameters in the form

$$\left. \begin{aligned} z_e &\equiv N_e \equiv (\frac{T_e}{T_0})^2 \equiv \frac{t_0}{\hbar / m_e c^2} \equiv \frac{l_0}{\hbar / m_e c} \equiv \frac{m_e}{m_x} \equiv \frac{a_e}{g_0} \equiv \cdots \equiv. \\ &\equiv (\partial c)^2 A_{S0}(m_e c / \hbar) \equiv \partial^2 c A_{S0}(m_e c^2 / \hbar) = 2.21(13) \times 10^{40}, \end{aligned} \right\} \tag{5.8a}$$

where the created temperature of electron since of the Big Bang $T_e$ is determined by

$$T_e \equiv \left(c\partial / \hbar_e\right)^{1/2} \equiv \left(\partial / t_e\right)^{1/2} = 4.113502\left(32\right)\times 10^{20}\,K. \qquad (5.8b)$$

(Because $T_e$ determined by specified values $\hbar_e$ and $t_e$, we have also $T_e^2 \hbar_e \equiv c\partial$ and $T_e^2 t_e \equiv \partial$). To obtain estimate on the cosmic acceleration $a_e$ at the electromagnetic time phase we may write

$$a_e \equiv g_0 \cdot z_e \equiv c^2 (m_e c / \hbar) = 2.32742099(3)\times 10^{29}\,m/s^2 \ . \qquad (5.8c)$$

Here the ratio $m_e / z_e$ corresponding to the mass of some unknown particle, must be

$$m_x = m_e / z_e = (\hbar / c) / l_0 \ . \qquad (5.8d)$$

Needs to note that, there the value of dimensionless large "magical" number (Davies 1982, Ch. 4) $z_e \equiv N_e \sim 10^{40}$ cannot be explained directly within the conventional hot Big Bang model. For this reason Carr and Rees (1979) points out that *"the task of deriving such a large pure number that $\sim 10^{40}$ from basic theory might seem a daunting one"*. However, according to relations (3.10) and (5.8a) it is should that this value is numbers of electrons in the Universe at the electromagnetic time moment of $t_e$ after the Big Bang.

Below we consider the electromagnetic, electroweak phases, the Planck, Trans-Planck and Over Trans-Planck epochs of the Universe. It is also possible should be hold for particles such as $W^\pm$ bosons and others. However, it appears to be some difficulties to do so in the context of proton and neutron, since these particles can not be considered as truly fundamental constants (see e.g., Tegmark et al. 2006; Fritzsch 2009) associated with their internal structure. Here we will address these questions in a future investigation.

### 5.3.1.2. Redshift in the electroweak phase

Here for the completeness we extend previous studies (Gasanalizade 2004, 2007) include a brief discussion on the redshift behaviour of the neutral boson $Z^0$ with the correct predicted mass of $m_{Z^0} = 91.1876(21)\,GeV/c^2$. Because region $100\text{-}1000\,GeV$ between electromagnetic phase and Planck epoch is one initial in our particle redshift investigations, for the *CERN's* high energy *Large Hadron Collider (LHC)* project it is very close to the experimental (14 TeV) bound (Ellis 2003).Then, the above energy is the line of demarcation between our local particle physics limitation and the energy in the early Universe after the onset of the Big Bang. However, it may be hoped that the energy regions that we wanted able to study, in both cases, have showed unexpected surprising results!

Our cosmological parameters of the $Z^0$ boson are determined on the basis of above value of mass recent published by Tegmark et al. (2006) and Mohr et al. (2008) derived from the new Particle Date Group data (Yao et al. 2006). We can then define the following accurate parameters relating to the mass $m_{Z^0}$:

1. The Compton wavelength of the boson should as an event horizon

$$\bar{\lambda}_{cZ^0} \equiv l_{Z^0} \equiv \hbar / m_{Z^0} c = 2.16397(50) \times 10^{-18} m . \qquad (5.9a)$$

2. This leads to the result that the creation moment of $Z^0$ bosons after the Big Bang is

$$t_{Z^0} \equiv \hbar / m_{Z^0} c^2 = 7.21823(17) 10^{-27} s , \qquad (5.9b)$$

3. The cosmic creation temperature of the $Z^0$ bosons (when *"it would acquire mass"*) is related to the $\bar{\lambda}_{cZ^0}$ and $t_{Z^0}$ by

$$T_{Z^0} \equiv (\partial c / \bar{\lambda}_{cZ^0})^{1/2} \equiv (\partial / t_{Z^0})^{1/2} = 1.73768(40) \times 10^{23} K . \qquad (5.9c)$$

This implies that the temperature $T_{Z^0}(\cos m)$ about of orders $10^8$ higher than *the threshold temperature* (Weinberg 1977) $T_{Z^0}(thr) = m_{Z^0} c^2 / k = 1.06 \times 10^{15} K$ . If this interpretation of $T_{Z^0}$ proves true (which it seems likely is no doubt), then experimental confirmation of above value $m_{Z^0}$ would rule out *CERN's* estimation of $T_{Z^0}(thr) \sim (10 - 100) TK$ . It should be clear that, the predicted as an initially cosmic creation temperature $T_{Z^0} \sim 10^{23} K$ , which dominated the Universe $\sim 10^{-26} \sec onds$ after the onset of the Big Bang, certainly lies beyond the realm of the present-day terrestrial laboratory measurements, since "*our experimental access to extremely early times is both in practice and in principle very limited*" (Bjorken 2004). It is conceived that, derived bound for cosmic creation temperature of $Z^0$ boson is significantly large than any obtainable by direct experiments of physical objects at the present time, in the *CERN's* high energy *LHC* project. This is may be a problem in further particle physic experiments. On the other hand, Sakharov and Hofer (2003) argued that "*the physical conditions that take place in our high energy colliders are very different from those that occurred in the early universe*". Then the correct explanation this issue deserves furthers both the theoretical and experimental investigations.

The redshift expression (or $N_{Z^0}$) for creation moment of $Z^0$ boson after the Big Bang is

$$\left.\begin{array}{l} z_{Z^0} \equiv N_{Z^0} \equiv \left(\dfrac{T_{Z^0}}{T_0}\right)^2 \equiv \dfrac{t_0}{t_{Z^0}} \equiv \dfrac{l_0}{l_{Z^0}} \equiv \dfrac{m_{Z^0}}{m_x} \equiv \dfrac{a_{Z^0}}{g_0} = \cdots \equiv \\[3mm] \equiv (\partial c)^2 A_{s0}(m_{Z^0}c/\hbar) = 3.95(23)\times 10^{45}. \end{array}\right\} \qquad (5.9d)$$

Here assuming $a_{Z^0}$ as a cosmic acceleration at the electroweak interaction phase, we obtain

$$a_{Z^0} \equiv g_0 \cdot z_{Z^0} \equiv c^2 (m_{Z^0}c/\hbar) = 4.15327(96)\times 10^{34}\, m/s^2. \qquad (5.9e)$$

As in the case of electromagnetic phase, Eq. (5.9d) also predicted the unknown mass

$$m_x = m_{Z^0}/z_{Z^0}. \qquad (5.9f)$$

What is $m_x$ in above equations? We again obtain surprising cosmological feature, when the ratio the particle mass to/redshift by the electron and the $Z^0$ boson give identical mass. This and other little one, is found also in case of following Planck and Trans –Planck redshift relations, which ultimately led to the identification of this with graviton mass $m_{Gr}$ and other with the string mass now $m_{Str}$. In order to compare this prediction of $m_{Gr}$ with the results from the Eq. (5.8d) and postulating that the Planck scale with the mass $m_{Pl} = (\hbar c/G)^{1/2} = 1.22\times 10^{19}\, GeV/c^2$ is not a fundamental scale (Arkani-Hamed et al.1998), we perform here a more rigorous calculations of the graviton mass on the framework of the laboratory and accelerator measurements through a comparisons between electromagnetic and $Z^0$ bosons phases redshifts. As a result, we find a correct lower limit for $m_{Gr}$ that compatible with the electron and boson masses, defined as

$$m_{Gr} = 4.11964(10)\times 10^{-71}\, kg = 2.31095(6)\times 10^{-35}\, eV/c^2. \qquad (5.10)$$

This is in excellent agreement with calculations from the Planck redshift relations shown below (Subsec. 5.3.3). Note that this implies also that two particles of different spins (the spin-1/2 electron, and a spin-1 $Z^0$ boson) are relates to the "cosmic particle" with a spin-2. The other good news is that predicted new cosmic particle (graviton), with masses forty orders of magnitude is smaller than the electron mass. Such a small mass could imply the existence of a new observable long-range force in nature in addition to quantum gravity, electromagnetism and electroweak interactions. On the other hand, this type of predictions suggest on existence of supersymmetrising gravity relation between bosons, fermions and gravitons (Salam 1982). Thus, although graviton mass has not yet been experimentally detected, this *"an incredibly small number"* (Carroll 2001) may be obtained directly by a laboratory data of particle physics with great precision!

An additional point to emphasize is that, we here determined cosmological redshift parameter $z_{Z^0} \equiv N_{Z^0} \sim 10^{45}$, which correspondents to the energy (and temperature) needed to the creation of the $Z^0$ boson in the early Universe and possible in the LHC experiments. But, in this case, what is the scalar Higgs boson mass? This boson is not observed in the experiments. Hawking (2008) believe that *"we won't find the Higgs"*. Nevertheless, it is conceived that *LHC* in *CERN* have generated this particle (Fritzsch 2009).

### 5.3.2. Redshift in the Planck epoch

Recently, problems of the early universe in the Planck era is considered by Bonnanno and Reuter (2002) in the modify standard isotropic flat Friedmann-Robertson-Walker (FRW) metric during the first few Planck times after the Big Bang. The main important assumption derived by Bonnanno and Reuter means that the dominant quantum corrections are correctly incorporated by substituting the time-dependence Newtonian gravitational constant as $G_0 \rightarrow G(t)$ and the Einstein cosmological constant as $\Lambda_0 \rightarrow \Lambda(t)$. Considering that high precision measurements give strong evidence that constant $G$ is time independent (see e.g. Sect.2.1, Will 2006 and their references), we assume that $G$ and also other FPC's starting from present-day laboratories $(z_0 < 1)$, till a new bound, corresponding to $z \mapsto \infty$, remains constant. However, following (Sahni and Starobinsky 2000; Carroll 2001; Bonnanno and Reuter 2002; Maia and Lima 2002; Padmanabhan 2003) it is assumed that the cosmological constant is time variable.

In the current cosmology, the Planck epoch is characterized by the four basic Planck parameters: $m_{Pl}, t_{Pl}, l_{Pl}$ and $T_{Pl}$ (Planck 1900). Here all parameters, except the Planck temperature $T_{Pl}$, are somewhat already known (Feynman et al.1999; Reynaud et al. 2008). From (Mohr and Taylor 2000, Mohr et al. 2008) it follows that

$$T_{Pl} = (\hbar c^5 / Gk^2)^{1/2} = 1.416785(71) \times 10^{32} K. \qquad (5.11a)$$

On the other hand, it has been argued that (Gasanalizade 2004) for correspondence with the Planck redshift, the Planck temperature derived from the (5.11a) could reduced an

$$T'_{Pl} = (\partial / t_{Pl})^{1/2} = \frac{(\hbar c^5 / G k^2)^{1/2}}{(4.965114231)^{1/2}} = 6.35828(32) \times 10^{31} K. \qquad (5.11b)$$

Here $T'_{Pl}$ denotes the reduced Planck temperature.

Using the Eq. (5.11b) the Planck redshift (also $N_{Pl}$) is read as (Gasanalizade 2007)

$$\left. \begin{array}{l} z_{Pl} \equiv N_{Pl} \equiv (\dfrac{T'_{Pl}}{T_0})^2 \equiv \dfrac{t_0}{t_{Pl}} \equiv \dfrac{l_0}{l_{Pl}} \equiv \dfrac{M_U}{m_{Pl}} \equiv \dfrac{m_{Pl}}{m_{Gr}} \equiv \\[3mm] \equiv \dfrac{m_{Gr}}{m_{Str}} \equiv \dfrac{a_{Pl}}{g_0} \equiv \cdots \equiv (\partial c)^2 A_{S0} (c^3 / \hbar G)^{1/2}, \end{array} \right\} \qquad (5.12a)$$

where $M_U$, $m_{Gr}$ and $m_{Str}$ are the masses of the Universe, graviton and string, respectively.

This leads to the predicted "*Planck acceleration*" $a_{Pl}$ given by

$$a_{Pl} = g_0 \cdot z_{Pl} = c^2 (c^3 / \hbar G)^{1/2} = 5.56074(28) \times 10^{51} m / s^2, \qquad (5.12b)$$

and

$$\left. \begin{array}{l} M_U \equiv M_\bullet(t) \cdot z(t) \equiv m_{Pl} \cdot z_{Pl} \equiv \cdots \equiv \\[2mm] \equiv (c^3 / G)(c \partial^2 A_{S0}) \equiv (c^2 / G) l_0 = 1.15(7) \times 10^{55} kg. \end{array} \right\} \qquad (5.12c)$$

Clearly, $z_{Pl} = \cdots = M_U / m_{Pl} = N_{Pl}$, can be identified also as the number of "*Planck particles*" at Planck moment in the early universe. Here, the mass of the Sun determined by Gundlach and Merkowitz (2000) as $M_S = 1.988435(27) \times 10^{30} kg$ tend to

$$M_U = 5.78(33) \times 10^{24} M_S. \qquad (5.12d)$$

This is some larger than a lower limit of its $M_U \geq 5.1 \times 10^{23} M_S$ obtained by Adler et al. (2006).

Let us also point that the result (relation 5) presented by Novello (2008) which "*interpret $M_g$ as the total mass of all existing gravitons in the observable universe*" and given by

$$M_g = \frac{\Lambda \cdot c^4}{G_N} \frac{1}{\sqrt{\Lambda^3}} \frac{1}{c^2}, \qquad (5.13)$$

can be identified with $M_U$, if is granted that $\Lambda = l_0^{-2} = \Lambda_0$ (here $G_N$ is Newton's gravitational constant). Thus, at the $t_{Pl} = (\hbar G / c^5)^{1/2} = 5.39124(27) \times 10^{-44} s$ (Mohr et al. 2008) we have

$$z_{Pl} \equiv N_{Pl} \equiv (\partial^2 c A_{S0})(c^5 / \hbar G)^{1/2} = 5.28(31) \times 10^{62}. \tag{5.14}$$

Throughout this paper, we will choose the "Planck redshift "as a supplementary fundamental "*cosmological quantum particle*" number of $N(t_{Pl}) \equiv N_{Pl}$, which can be interpreted as the number of the black hole with size of Planck mass $m_{Pl}$ (Carr and Rees 1979). The exact value of $N_{Pl}$ must be determined as a function of the Planck temperature $T_{Pl}^{'}$ from Eq. (5.11b) and CMB temperature $T_0 = 2.725(1) K$ (Fixsen and Mather 2002) as

$$z_{Pl} \equiv N_{Pl} \equiv (T_{Pl}^{'} / T_0)^2 = 5.444_4(2) \times 10^{62} . \tag{5.15}$$

For the lower and upper limits on $z$, we have $z_0 \le z \le z_{Pl}$, or $1 \le z \le 5.28(31) \times 10^{62}$.

Note that, in fact Planck redshift $z_{Pl}$ about $10^{62}$ in scientific literature is widely treated as the *60e* or, *61e- foldings (i.e., as a factor $e^{60} - e^{61}$ ;* see e.g., Buosso 2006).

As is noted above Eq. (5.12a) shows that *cosmological particles $m_{Pl}$*, $m_{Gr}$ and $m_{Str}$ characterized by a constant mass may be evolves according to

$$M_U : m_{Pl} : m_{Gr} : m_{Str} \equiv z_{Pl} \equiv N_{Pl} . \tag{5.16}$$

Another important consequences of *cosmological particles* would be their originate time. Since $m_{Pl} >> m_{Gr} >> m_{Str}$ the situation is different. This means that $t_{Str} << t_{Gr} << t_{Pl}$ after the Big Bang. We discuss in more detail the evolution of the universe beyond of $t_{Pl}$ in following sections, where is taken $t_{Gr} \equiv t_{TPl}$ and $t_{Str} \equiv t_{Crit}$ .

It is obvious from relation (5.12c) and special relativity that

$$M_U c^2 = 1.03(6) \times 10^{72} J = 6.45(37) \times 10^{90} eV , \tag{5.17}$$

which is four order of magnitude larger than the value estimated by Davies (1985, Ch.12) and is a 20 times more than this determined in (Misner, et al. 1973). Other an alternative close estimation for the mass of the observable universe suggested by Carr and Rees (1979)

$$M_U \sim c^3 t_0 G^{-1} H_0^2 \sim c^3 t_0^2 G^{-1} \sim \alpha_G^{-1} m_p^2 \big/ m_e \ , \tag{5.18}$$

where $m_p$ is proton mass and $\alpha_G = G m_p^2 / \hbar c = 5.91 \times 10^{-39}$ is "*the gravitational fine structure constant*". Then, in the case of (5.18) we have $M_U \sim 9 \times 10^{54} kg$ $\sim 5.05 \times 10^{90} eV / c^2$, which is a factor of 1.28 less than value of $M_U$ from (5.17), i.e., predicted $M_U c^2$ is larger than all the present higher bound on the mass of the Universe.

### 5.3.3. Redshift in a Trans-Planck epoch

First we outline relations between the cosmological constant $\Lambda_0$ and Planck parameter of $l_{Pl}$. The bound on $(\Lambda_0 l_{Pl}^2)^{-1}$ and $\Lambda_0 l_{Pl}^2$ arise from the square of the Planck redshift $z_{Pl}^2$ or – equivalently – from the inverse value $z_{Pl}^{-2}$, in Trans-Planck (*T-Pl*) time prior to the Big Bang, when the Universe was extremely smaller than, in Planck time $t_{Pl}$. In particular the quadratic Planck redshift $z_{Pl}^2$ can be interpreted as a Trans-Planck cosmological redshift (also $N^2(t_{Pl}) \equiv N_{Pl}^2 \equiv N_{TPl} \equiv N_{Gr}$) defined as

$$\left. \begin{aligned} z_{TPl} \equiv z_{Pl}^2 \equiv N_{Gr} &\equiv (\frac{T_{TPl}}{T_0})^2 \equiv \frac{t_0}{t_{TPl}} \equiv \frac{l_0}{l_{TPl}} \equiv \frac{M_U}{m_{Gr}} \equiv \cdots \equiv \\ &\equiv \frac{a_{TPl}}{g_0} \equiv \frac{u_\nu(t_{Pl})}{u_\nu(t_0)} \equiv \cdots = 2.79(30) \times 10^{125}, \end{aligned} \right\} \tag{5.19a}$$

where

$$T_{TPl} \equiv (T_{Pl}')^2 / T_0 = 1.46(8) \times 10^{63} K \ , \tag{5.19b}$$

$$t_{TPl} \equiv t_{Pl}^2 / t_0 \equiv \hbar / M_U c^2 = 1.02(6) \times 10^{-106} s \ , \tag{5.19c}$$

$$l_{TPl} \equiv l_{Pl}^2 / l_0 \equiv \hbar / M_U c = 3.06(18) \times 10^{-98} m \ , \tag{5.19d}$$

$$m_{Gr} \equiv m_{Pl}^2 / M_U = 4.11(24) \times 10^{-71} kg \ , \tag{5.19e}$$

$$a_{TPl} \equiv a_{Pl}^2 / g_0 \equiv 2.94(17) \times 10^{114} m / s^2 \ . \tag{5.19f}$$

Then, from the Eq. (5.19a) we have $(\Lambda_0 l_{Pl}^2)^{-1} \equiv (l_0 / l_{TPl}) \equiv (l_0 / l_{Pl})^2$. On the other hand, $M_U$ can be written as a function $z_{Pl}$ and masses of "cosmological particles" as

$$M_U \equiv m_{Gr} z_{Pl}^2 \equiv m_{Gr} (m_{Pl} / m_{Gr})^2 . \tag{5.19g}$$

Here we notice the similarities between $(m_{Pl} / m_{Gr})^2$ and provided by Reynaud et al (2008) ratio $(m / m_{Pl})^2$. These authors have devoted attention to estimate a role of the Planck mass in the definition of a borderline between microscopic and macroscopic masses.

In that case, one can evident from (5.19g). In particular, it is clearly leading to $m_{Gr} = m_{Pl}^2 / M_U$, which we will briefly discuss in Sect.6.3 [see also (5.21b), (5.21c)].

*5.3.4 Redshift in an Over Trans-Planck epoch*

The maximum (threshold) relativistic temperature conformity with the total mass of the Universe can be determined by relation (see e. g. Weinberg 1977, Ch. 4)

$$T_U = M_U c^2 / k . \tag{5.20a}$$

On the other hand, this limits in fact a little less [see, Eq. (5.11b)], so $T_U$ will reduce to

$$T_{Crit} = T_U^{'} = \frac{M_U c^2 / k}{(4.965114231)^{1/2}} = (\partial c^3 / G m_{Str})^{1/2} = 3.36(19) \times 10^{94} K . \tag{5.20b}$$

This would imply the existence of an *absolute maximum temperature* in history of the Universe. Extrapolating into the past, in the state, when the Universe originated in a largest temperature $T_{Crit}$, we obtain *the lowest initial time* corresponding to creation of the Universe

$$t_{Crit} \equiv \partial / T_{Crit}^2 \equiv G m_{Str} / c^3 = 1.93(11) \times 10^{-169} s . \tag{5.20c}$$

This is combined with the horizon $l_{Crit}$ as

$$l_{Crit} \equiv c \cdot t_{Crit} \equiv \partial c / T_{Crit}^2 \equiv G m_{Str} / c^2 = 5.79(33) \times 10^{-161} m . \tag{5.20d}$$

This is *initial event horizon* of the universe at moment $t_{Crit}$ after the onset of the Big Bang. Then, $l_{Crit}$ is a minimum permissible meaningful size at very early structure of the Universe, below which the horizon disappears. This is extremely important parameter in cosmology predicted by string theory (Green 1999, Ch.14). On the other hand, $t_{Crit} = t_{Str}$ by Eq. (5.20c) is the smallest meaningful time moment in the evolution of the Universe, when one has attained this limit value of size $l_{Crit}$. Here a primary consideration is the possibility of avoiding the initial singularity (see, e.g. Penrose 1969) and replacing this. We obtained as the best-fit that the Big Bang has arise "much latter" at $0 < t \leq t_{Crit} \sim 10^{-169} s$, rather than in a naked classical singularity $t = 0$, corresponding to the Friedmann-Lemaitre cosmology. Then, at a later stage the singularity is reduced to a predicted discrete time moment of $t_{Crit} \cong 2 \times 10^{-169} s$. In fact, this critical moment can possible be identified with the beginning of the Universe time.

Finally, we need theoretical paradigm for the inflationary Big Bang model of the Universe. Many scientists just believe that accelerated expansion ("inflation") of the Universe immediately is entered a phase of $t \sim 10^{-35} s$ since the Big Bang (Guth 1981; Linde 1982; Albrecht and Steinhardt 1982; Liddle and Lyth 2000; Guth and Kaiser 2005; Mukhanov 2005; Baumann and Peiris 2008, and references therein). The above equations lead to initial acceleration of the Universe defined as

$$a_{OTPl} \equiv a_{crit} \equiv c / t_{crit} \equiv (c / \partial) T_{crit}^2 = 1.553(88) \times 10^{177} \, m / s^2 \, . \qquad (5.20e)$$

Moreover, the critical condition "*at the extremely very early epoch of the development of the Universe, when even the conception of matter in usual sense did not exist yet*" (Sakharov and Hofer 2003). It is straightforward describes by the Over Trans-Planck (*OTPl*) epoch at very highest redshifts (also $N_{Pl}^3 \equiv N_{OTPl} \equiv N_{Crit} \equiv N_{Str}$), given by

$$\left. \begin{aligned} z_{OTPl} \equiv z_{Crit} &\equiv z_{Pl}^3 \equiv N_{Str} \equiv (\frac{T_{Crit}}{T_0})^2 \equiv \frac{t_0}{t_{Crit}} \equiv \frac{l_0}{l_{Crit}} \equiv \frac{M_U}{m_{Str}} \equiv \\ &\equiv \frac{a_{Crit}}{g_0} \equiv \cdots = 1.48(26) \times 10^{188}, \end{aligned} \right\} \qquad (5.21a)$$

where $N_{Str}$ is the number of string in the observable Universe, and $\sum_{1}^{N} m_{Str} \equiv M_U \equiv z_{Crit} \cdot m_{Str} \equiv N_{Str} \cdot m_{Str}$. Eventually we shall have the following sequence of identities

$$\frac{M_U}{m_{Pl}} \equiv \frac{m_{Pl}}{m_{Gr}} \equiv \frac{m_{Gr}}{m_{Str}} \equiv \frac{N_{Gr}}{N_{Pl}} \equiv \frac{N_{Str}}{N_{Gr}} \equiv \cdots \equiv z_{Pl} \equiv N_{Pl} \, , \qquad (5.21b)$$

and

$$\frac{M_U}{m_{Gr}} \equiv \frac{m_{Pl}}{m_{Str}} \equiv \frac{N_{Str}}{N_{Pl}} \equiv \cdots \equiv z_{Pl}^2 \equiv N_{Gr}. \tag{5.21c}$$

These ratios are the hierarchy in the mass scale of cosmological "particles".

### 5.4. The Universe with the Constant Mass and Energy

It could be argued that there $M_U$ is a total mass of the Universe given by Eq. (5.11e). A crucial element in the relation (5.11e) is the general acceptance of $M_\bullet(t) \cdot z(t) = (c^2/G)l_0$, which produce the paradox, namely, which of these parameters is constant or as variable? The Standard Model provided a reason to accept constants $c$ and $G$ (see, e.g., Sect. 2), it also ratio $c^2/G$ ($\equiv m_{Pl}/l_{Pl}$) without the change as physical constituents the underlying laws of current physics (see e.g., Planck 1900; Treder 1979). *"Consequently, a time dependent speed of light is equivalent to a time dependent gravitational constant"* (Lammerzahl et al. 2006). Some physicists believe that *"if one finds that the fundamental constants are changing in time, then they are not just numbers, but dynamical quantities which change according to some deeper laws that we have to understand. These laws would be truly fundamental and may even point the way to a unified theory including gravity"* (Fritzsch 2009).

In a today Universe model, where $z(t) \equiv z_{Pl}^0 \equiv z_0(t_0) \equiv z_0 \equiv 1$ and the total mass $M_U$ is fixed, the value of $l_0$ also becomes constant as *the Einstein radius of the universe* at the present epoch. In other case, if accept the $z = z(t)$, at $l_0 = Const.$, the mass $M_U$ of an Universe also varies, which is the risk of violating the first law of thermodynamics (Davies et al. 2002; Carlip 2003). However, it could be argued that in process of the isotropic and homogeneous expansion of the early Universe the product $M_\bullet(t) \cdot z(t)$ of two cosmological parameters remains constant all times, leading to a generalization of the first law of thermodynamics. Thus, starting assumptions was that, $M_U$ remain constant as $z(t)$ and $M_\bullet(t)$ changes. The Planck redshift model predict a four period of time corresponding to a steep redshift increase in the early Universe evolution. On the other hand, there are (rather exactly defined) upper and lower limits to $M_\bullet(t)$, but within the range, there is an upper limit to $M_U$ and lower limit (as already discussed) to string mass $m_{Str}$.

Let us now summarize these parameters. The specific theoretical model list with discrete time steps of $\Delta t \sim 10^{-62} s$ (from the present epoch) is included to the table 2.

**Table2**. The mass of cosmological particles expressed in terms of Plank redshift steps.

| Phase | $t$ | $z(t) \equiv N^n$ | $M_*$ | $M_U$ |
|---|---|---|---|---|
| I | $t_{OTPl}$ | $z_{PL}^3 \equiv N_{Pl}^3$ | $m_{Str}$ | $m_{Str} \cdot z_{Pl}^3 \equiv m_{Str} \cdot N_{Pl}^3$ |
| II | $t_{TPl}$ | $z_{Pl}^2 \equiv N_{Pl}^2$ | $m_{Gr}$ | $m_{Gr} \cdot z_{Pl}^2 \equiv m_{Gr} \cdot N_{PL}^2$ |
| III | $t_{Pl}$ | $z_{Pl} \equiv N_{PL}$ | $m_{Pl}$ | $m_{Pl} \cdot z_{Pl} \equiv m_{Pl} \cdot N_{Pl}$ |
| IV | $t_0$ | $z_o = 1$ | $M_U$ | $(c^2/G)l_0 \equiv M_U \cdot N_{Pl}^0$ |

We see from above equations and table 2 that in the temporal expansion of the Universe there is no violation of the generalized first law of thermodynamics. Since $T_{OTPl} >> T_{TPl} >> T_{Pl} >> T_0$ it would seem that the generalized second law of thermodynamics also is not violated. It is thus clear that this connection between thermodynamics and constant ratio $c^2/G$ obtained under the assumption that modified general relativity description of gravity also is correct at all times up to $t_{OTPl} \sim 10^{-169} s$. Other a striking fact is that the time (temperature) dependence of the Freundlich redshift beginning at current epoch of the Universe and ending in beyond the Planck epoch has showed of multiple discrete space-time structures (see also, Padmanabhan 2003, p.66). This may be done as a replace of continuum by a quantization (Sakharov and Hofer 2003; Kimberly and Magueijo 2005).

The expansion history of the early Universe vs. redshift, time and temperature for illustrative purpose is shown in figure 1. Here we present key events of the time history of the very early Universe at a four epochs of cosmic expansion, with a very significant discrete step that the ratios $z_{OTPl} : z_{TPl} : z_{Pl} : z_0 \equiv N_{Str} : N_{Gr} : N_{Pl} : N_0$ are of order about $10^{62}$. On the other hand, these parameters just as showed that (see also, Eqs. (5.19a) - (5.19t) and (5.20c)) the background cosmological particles-gravitons and strings originates (creates) at the Trans-Planck times $t_{TPl} \sim 10^{-106} s$, and Over Trans-Planck times at $t_{OTPl} \sim 10^{-169} s$, respectively. However, this interpretation leads to the apparently inescapable conclusion (see also Subsec.5.3.4) that the Universe has a beginning of time $t_{OTPl}$, while in this (the strings) case, the quantizing gravity by the concepts of the Heisenberg uncertainty relation losing their applicability. We infer from this that quantize gravity will be a valid over the times range, corresponding to $t \geq t_{TPl}$.

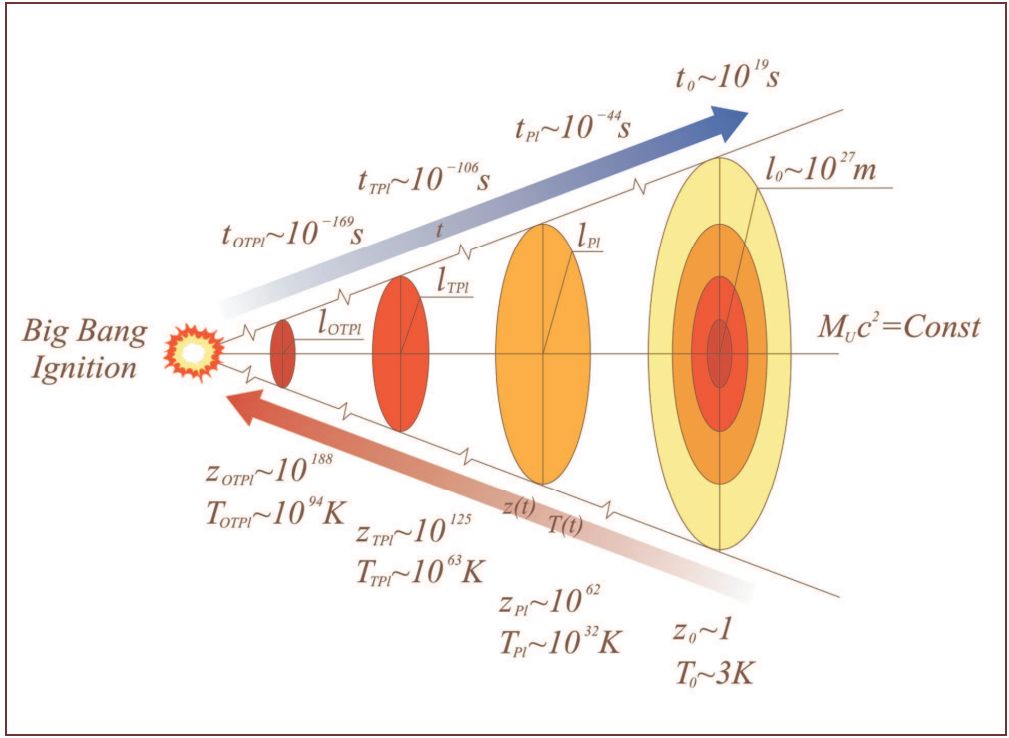

Figure 1. The "$10^{62}$-folding" extension model of the early Universe following of Big Bang ignition, over a wide range in time and temperature scale, illustrating the some points discussed above. There the Planckian redshift is intended to do, a four–steps of "$10^{62}$- folding" process. One sees that, the "arrow of time" is started out from $t_{Crit} \sim 10^{-169}$ seconds, when the Universe's horizon size should be minimal and equal to $l_{Crit} \sim 10^{-161}$ meter, whereas the both of "arrows of temperature and redshift" are aimed away from us to the past.

## 5.5. The time dependence of the cosmological acceleration and on constancy of the velocity of light

In the "synthesis" context, the cosmological accelerations are constant on epoch scales and decay as $\sim a^{-62}$ over the four cosmic redshifts steps, and thus

$$a_{OTPl} \gg a_{TPl} \gg a_{Pl} \gg \cdots \gg g_0(t_0) \sim 10^{-11} m/s^2 . \qquad (5.22)$$

The cosmic time is related to the cosmological acceleration by (see Sec. 5)

$$\left.\begin{array}{l} a_{OTPl} \cdot t_{OTPl} \equiv a_{TPl} \cdot t_{TPl} \equiv a_{Pl} \cdot t_{Pl} \equiv \cdots \equiv \\ \equiv a_{Z^0} \cdot t_{Z^0} \equiv a_e \cdot t_e \equiv g_0 \cdot t_0 \equiv c \equiv Const. \end{array}\right\} \qquad (5.23)$$

Note that this scenario leads to a time depended decrease of the cosmological acceleration during the cosmic expansion of the Universe. We then have result that the deceleration parameter $q_0$ can be interpreted as $q_0 > 0$. This prediction is incompatible with the cosmological acceleration which has been pursued in initial supernova observations (Riess et al. 1998; Perlmutter et al. 1999) (see also, Percival et al 2001; Hu 2002; Sievers et al. 2003; Riess et al. 2004; Moffat 2005; Copeland et al. 2006 and their references), where the deceleration parameter $q_0 < 0$ corresponding to an accelerating Universe and may be masking the true situation. This point of view has been taken also by Carroll (2003), who argues that *"The acceleration of the Universe presents us with mysteries and opportunities. The fact that this behaviour is so puzzling is a sign that there is something fundamental we don't understand. We don't even know whether our misunderstanding originates with gravity as described by general relativity, with some source of dynamical or constant dark energy, or with the structure of the universe on ultra-large scales"*. That the expected acceleration of the expansion of the Universe is profound mysteries and questionable are considered also in the other studies (see, e.g., Wetterich 2003; Geshnizjani et al. 2005; Flanagan 2005; Hirata and Seljak 2005; Apostolopoulos et al. 2006; Wood-Vasey et al. 2007, and their references). In any case, on the subject of coincidences by Eq. (5.23) we have concluded that *for all discrete steps of the expansion of the Universe the product of cosmological acceleration after the Big Bang on a time in this moment is kept constant, which is exactly equal to speed of light in vacuum.*

### 5.5.1. Other explanation

Eq. (5.23) means also that there exists other cosmological law, where acceleration $a(t)$ after the Big Bang is rigidly bound with the squared temperature at different epochs given by

$$\left.\begin{array}{l} a_{OTPl} / T_{OTPl}^2 \equiv \cdots \equiv g_0 / T_0^2 \equiv c / \partial = \\ = 1.37548(76) \times 10^{-12} ms^{-2} K^{-2} = Const. \end{array}\right\} \qquad (5.24)$$

For example, at *CMB* temperature $T_0 = 2.725(1) K$ (Fixsen and Mather 2002) today epoch this model predicts a critical minimal value of $g_0 = 1.0214(4) \times 10^{-11} m/s^2$ (see also table 3), then is ruled out an accelerating Universe. This cosmological approach once again tells us about other fundamental issues such as *the constancy of the velocity of light in all phases of the cosmic expansion of the Universe*, given by Eq. (5.23).

## 6. The mass formula for a graviton and strings

It is generally believed that, the graviton is the quant of gravitation (see e.g. Schwarz, 1987). However, it is not enters into Standard Model (SM) of particle physics. The occurrence of the graviton mass no envisaged by general relativity (Weinberg, 1972; Sivaram, 1999). So far, well know the graviton have the spin 2 (i. e., $4 \times (\hbar / 2)$). The Particle Data Group (PDG 1996) noted that an astrophysical experimental limit of graviton mass is (see also, Visser 1998)

$$m_{Gr} < 2 \times 10^{-29} \, eV / c^2 \approx 2 \times 10^{-38} m_{nucleon} \,, \tag{6.1}$$

that corresponds to a Compton wavelength of

$$\lambda_{Gr} = \frac{h}{m_{Gr} c} > 6 \times 10^{22} m \,. \tag{6.2}$$

### 6.1. Graviton mass from the early research

Astronomical work on the graviton mass goes back to Savchenko (1949), which derived $m_{Gr} \sim 10^{-69} kg$. Gomide (1963) for a density of the $10^{-28} kg \cdot m^{-3}$ "empty" Universe consider revising fond that $m_{Gr} \sim 10^{-68} kg$. Kurdgelaidze (1964) found that at the value of Hubble constant $H_0 = 2.5 \times 10^{-18} s$, $m_{Gr} = 4 \times 10^{-69} kg$. Hiida and Yamaguchi (1965) suggest that the analysis of the dynamics of clusters of galaxies could yield a limit on the graviton mass as low as $5 \times 10^{-65} kg$. Goldhaber and Nieto (1974), by recalling that the best graviton mass limit comes from a larger distances in system of galaxies and using the distance $r = 580 kpc$ yields a limit for $m_{Gr}$ of $m_{Gr} \leq 2 \times 10^{-65} kg = 5.6 \times 10^{-27} m^{-1} = 1.1 \times 10^{-29} eV / c^2$.

### 6.2. Today situation

Gershtein et al. (2004), using the relativistic field theory of gravity (RTG) and measured value of $\Omega_{tot}$ (the total cosmological density) obtained the upper limit on the graviton mass with 95% CL: $m_{Gr} \leq 1.6 \times 10^{-69} kg$, within the $1\sigma$ its probable value is derived as $m_{Gr} = 1.3 \times 10^{-69} kg$. Wesson (2004) by using a phenomenological approach has found quantized mass according to the rule $m = (n\hbar / c)(\Lambda / 3)^{1/2}$, from which in case $n = 1$,

"minimal" mass defined as $m_P = (\hbar/c)(\Lambda/3)^{1/2} \approx 3.5 \times 10^{-69} kg$. We have shown in Sect.5 (using detailed simulations) that some predicted new and presumable very light elementary particles, with the masses of the order of $10^{-71} kg$, and $10^{-134} kg$, respectively, can be appertain to a graviton and it is possible to strings.

Will (1998) believe that the mass of the graviton can be estimated using the bound of graviton Compton wavelength $\lambdabar_g$ which is of the order of $6 \times 10^{15} m$ for the ground-based laser-interferometers of the LIGO/VIRGO type, and $\lambdabar_g > 6 \times 10^{19} m$ for the proposed Laser Interferometer Space Antenna (LISA). However, from the observation of CMB anisotropies allows one to indirectly constrain the primordial background of gravitational waves with frequencies $\sim 10^{-18}$ around to $10^{-16} Hz$ (Kamionkowski and Kosowsky 1999; Seto and Corray 2006). On the other hand, attempts to detect primordial waves at frequencies around $10^{-7} Hz$ do not be able because their amplitude with today's technology is expected to be small (Schwarz 2003). Of course there may be other ways to of extensive searches the "massive graviton" detection. Recent estimates suggest that with the LISA which to be launched in 2018, or thereabouts it will try to detect via gravitational waves (GWs) extraordinary cosmic events, such as the birth of galaxies and the death of black holes (Bell 2008). In addition, by a comparison of the orbital phase related to an optical eclipsing light curve of close white dwarf binaries (CWDBs) with the phase determined from LISA data will allow a constraint on the graviton mass. Then, using a sample of CWDBs detectable with the LISA and the associated optical light curve data related to binary eclipses the upper limit for graviton mass is determined at the level of $\sim 6 \times 10^{-24} eV$ (Cooray and Seto 2004) if one assumes a gravitational propagation speed is less than of the speed of the electromagnetic interaction. The recent analysis assumes (Will 2006) that for super-massive binary black holes (SMBBH) with the masses $M_{SMBBH} \sim 10^4$ to $10^7 M_S$, observations by LISA can be estimate the bound on massive graviton Compton wavelength as $\lambdabar_g > 10^{22} m$ identical with those of Eq. (6.2). This value $\lambdabar_g$ is appropriate to bound on mass as $m_g < 10^{-66} kg$. This bound is almost five orders of magnitude greater than the prediction by a cosmological model (see, Eq. (5.10)). For a widely details of this approach, see the recent valuable research of Arun and Will 2009, and references therein.

In the next subsection we presented predicted results in detail for the some case.

### 6.3. Graviton mass from various phases

1. In the case of the Planck and Trans-Planck redshifts we shall have the sequence

$$m_{Gr} \equiv m_{Pl} / z_{Pl} \equiv M_U / z_{Pl}^2 \equiv m_{Pl}^2 / M_U. \tag{6.3}$$

Here the value $m_{Gr} = m_{Pl}^2 / M_U$ can be interpreted also as a results of the so-called "*seesaw mechanism*", which has been applied by Gell-Mann et al. (1979; see also, Arkani-Hamed et al. 2000) for determination of neutrino mass. However, few cosmologists and particle physicists believe (see e.g. Cline 1988) that *seesaw mechanism* is a complete *ad hoc* assumption. Thus in such a case, the *"seesaw mechanism"* for determination of graviton mass have acquired of theoretical basis.

2. The correct value of the electron-graviton mass ratio $m_e / m_{Gr}$ in the electromagnetic phase of the Universe is expressed as

$$m_e / m_{Gr} \equiv z_{el.ep} \equiv (\partial c)^2 A_{S0} (m_e c / \hbar) = 2.21(13) \times 10^{40}, \qquad (6.4)$$

from which graviton mass at once is (see, also, Eq. (5.10))

$$\left. \begin{array}{l} m_{Gr} \equiv (\hbar / c) / l_0 \equiv (\hbar / c) \Lambda_0^{1/2} \equiv (\hbar / \partial c^2) T_0^2 = \\ = 4.11(24) \times 10^{-71} kg = 2.31(13) \times 10^{-35} eV / c^2. \end{array} \right\} \qquad (6.5)$$

For $T_0(CMB) = 2.725(1) K$ (Fixsen and Mather 2002) we have

$$m_{Gr} \equiv (\hbar / \partial c^2) T_0^2 = 3.9976(29) \times 10^{-71} kg = 2.2425(16) \times 10^{-35} eV / c^2. \quad (6.6)$$

This corresponds to the graviton Compton wavelength

$$\lambdabar_{Gr} \equiv l_0 \equiv \hbar / m_{Gr} c = 8.799(6) \times 10^{27} m \quad . \qquad (6.7)$$

Here $m_{Gr}$ from (6.6) is 32 times less than those obtained in (Gershtein et al. 2004).

Similar to Eq.(6.5) relation for the graviton mass recently obtained also by Novello (2008), which writes as

$$m_g = (\hbar / c) \sqrt{\Lambda} , \text{ with } m_g \approx \sqrt{\Lambda} , \qquad (6.8)$$

where $\Lambda$ "*is treated as a fundamental constant related to the gravitational interaction*". If denote $\Lambda = \Lambda_0 = l_0^{-2}$ (as this is identified in Sec. 5.4), we should have $m_g \equiv m_{Gr}$.

Recent information on photon and graviton mass limits and associated Compton wavelengths with a great citation (270 references) are summarized by veterans of this problem Goldhaber and Nieto (2009). In particular, for a minimal of graviton mass and their maximum Compton wavelength derived results by the Dvali-Gabadadze-Porrati (DGP) model (Dvali et al. 2000; Dvali et al. 2003) are respectively

$$m_{Gr} \leq \sim 10^{-69} \, kg \equiv 6 \times 10^{-34} \, eV/c^2, \qquad (6.9)$$

and

$$\lambdabar_{Gr} \geq 3 \times 10^{26} \, m, \qquad (6.10)$$

which are a some differ from the one obtained by (6.6) and (6.7).

On the other hand, the combine of the electron and the $Z^0$ boson masses with the graviton masses by relations (5.8a), (5.9d) and (6.4) may be treated an uniquely attempt of the integration of the particles with the various spins (Hawking 1988).

3. Finally, from relation (5.15) we obtain a limiting smallest value of minimum mass $m_{Str}$ in the system a family of "cosmological particles". This value may be taken as the absolute minimal mass an appertain to the string, which related to the set of above constant cosmological particles masses in the form

$$\left. \begin{aligned} m_{Str} &\equiv m_{Gr}/z_{Pl} \equiv m_{Gr}^2/m_{Pl} \equiv (\hbar/c)\Lambda_0 l_{Pl} \equiv (m_{Pl}/M_U)m_{Gr} \equiv \cdots = \\ &= 7.78(45) \times 10^{-134} \, kg = 4.37(25) \times 10^{-98} \, eV/c^2. \end{aligned} \right\} \qquad (6.11)$$

The latter is much smaller than "the absolute minimal mass" obtained in (Bohmer and Harko 2006)

$$M_{\min} = \frac{\Lambda c^2}{12G} l_{Pl}^3, \text{ where } M_{\min} \approx 1.4 \times 10^{-124} \, kg \approx 7.9 \times 10^{-95} \, eV. \qquad (6.12)$$

However, it is clearly seen that, the value of $M_{\min}$ at least of the order $10^{10}$ of magnitude large than $m_{Str}$ from (6.11). Then $m_{Str}$ from (6.11) can be considered as a preliminary estimation of "*minimal cosmological mass*" known at present.

## 7. The cosmological constant problem

Now let us return to the issue of the cosmological constant problem. Cosmological constant was introduced by Einstein (1917) to solve the discrepancy between an apparently static Universe and the dynamic cosmology of general relativity. However, after the known works of Friedmann (1922, 1924) and the Hubble's discovery of the expansion of universe in 1929 Einstein discarded the cosmological constant in favour of expanding model without once for all. This was restored in the late 1960s by Sandage (1961), Petrosian et al (1967), Petrosian (1974) and et al. Reviews the broad history on a positive cosmological constant has been investigated by Sahni and Starobinsky (2000). The conceptual and fundamental aspects of this problem is well-described by Padmanabhan (2003) and Peebles and Ratra (2003).In spirit of this analysis we seek to use value of $\Lambda_0$ that for today universe leads to (see, e. g., Petrosian 1974; Islam 1992; Sahni and Starobinsky 2000; Padmanabhan 2003)

$$\Lambda_0 \equiv l_0^{-2} \equiv \frac{8\pi G}{c^4} \cdot u_\nu(t_0) \equiv \frac{8\pi G}{c^2} \cdot \rho_\Lambda(t_0), \tag{7.1}$$

where $\Lambda_0$ is the today value of cosmological constant. In atomic units $\Lambda_0 = (m_{Gr} c / \hbar)^2$, then $\Lambda_0$ is inversely proportional to the square of Compton wavelength of graviton, and generalized to include graviton mass, which is constant in time. In this case, $l_0$ is "*the radius of closed spherical space today universe*" in near its maximum of expansion, $u_\nu(t_0) = \rho_\Lambda(t_0) \cdot c^2$ is the minimal vacuum energy density of pressure-less matter (Padmanabhan 2003) and $\rho_\Lambda(t_0)$ is the minimal vacuum matter density in the today Universe. We may express the today vacuum energy density in terms of a cosmological constant as follows

$$\left. \begin{aligned} u_\nu(t_0) \equiv \rho_\Lambda(t_0) \cdot c^2 &\equiv \frac{c^4 \Lambda_0}{8\pi G} \equiv (c^4 / 8\pi G)(m_{Gr} c / \hbar)^2 \equiv \cdots = \\ &= 6.61(40) \times 10^{-14} J \cdot m^{-3} = 4.12(24) \times 10^6 eV \cdot m^{-3}. \end{aligned} \right\} \tag{7.2}$$

Similarly, the minimal vacuum matter density $\rho_\Lambda(t_0)$ of the today universe is

$$\rho_\Lambda(t_0) \equiv \frac{c^2 \Lambda_0}{8\pi G} \equiv (c^2 / 8\pi G)(m_{Gr} c / \hbar)^2 = 7.35(43) \times 10^{-31} kg \cdot m^{-3}. \tag{7.3}$$

Alternatively, for example, one could define graviton mass $m_{Gr}$ in terms of the today vacuum energy density and matter density. We find

$$m_{Gr} \equiv (\hbar / c^3)[8\pi G \cdot u_\nu(t_0)]^{1/2} \equiv (\hbar / c^2)[8\pi G \cdot \rho_\Lambda(t_0)]^{1/2}. \tag{7.4}$$

### 7.1. Essence of the "Weinberg paradox"

Note that the cosmological constant is the many-sided parameter varying in space-time. Consider now this variation. In particular, this interpolates between today epoch of universe value $\Lambda(t \xrightarrow{t} t_0) = \Lambda_0 = l_0^{-2}$, the Planck epoch as $\Lambda(t \longrightarrow t_{Pl}) = \Lambda_{Pl} = l_{Pl}^{-2}$, respectively. Then cosmological constant is different in the two epochs (see also, Sahni and Starobinsky 2000; Caroll 2001). Furthermore, in relativistic particle physics models $\Lambda$ is inversely proportional to the square of the Compton wavelengths of particle. For the Planck epoch we have

$$\Lambda_{Pl} \equiv \Lambda(t_{Pl}) \equiv (\hbar / m_{Pl}c)^{-2} \equiv \frac{8\pi G}{c^4} u_\nu(t_{Pl}) \ . \tag{7.5}$$

So that Eq. (7.5) for the energy density at the Planck scale may be rearranged as follows

$$\left.\begin{aligned} u_\nu(t_{Pl}) &\equiv \rho_\Lambda(t_{Pl}) \cdot c^2 \equiv (c^4 / 8\pi G)(m_{Pl}c / \hbar)^2 \equiv \frac{c^4 \Lambda_{Pl}}{8\pi G} \equiv \cdots = \\ &= 1.80(19) \times 10^{122} J \cdot m^{-3} = 1.15(12) \times 10^{132} eV \cdot m^{-3}. \end{aligned}\right\} \tag{7.6}$$

It is interesting to compare upper Planck limits of vacuum matter density $\rho_\Lambda(t_{Pl})$ and the vacuum energy density $u_\nu(t_{Pl})$ from Eq.(7.6), with the complemented by lower today limits of $\rho_\Lambda(t_0)$ and $u_\nu(t_0)$, from Eq. (7.2) and (7.3), respectively. This combination, however, leads to dramatically intriguing relationship (*"the Weinberg paradox"*). This can be carried out as follows

$$\rho_\Lambda(t_0) = 3.6 \times 10^{-126} \rho_\Lambda(t_{Pl}) \ \text{ and } \ u_\nu(t_0) = 3.6 \times 10^{-126} u_\nu(t_{Pl}) \,, \tag{7.7}$$

or

$$\Lambda_{Pl} / \Lambda_0 \equiv \rho_\Lambda(t_{Pl}) / \rho_\Lambda(t_0) \equiv u_\nu(t_{Pl}) / u_\nu(t_0) \equiv c^3 / \hbar G \Lambda_0 = 2.8 \times 10^{125}, \tag{7.8}$$

which Weinberg (1993, p.179) named as *"the worst failure of an order-of-magnitude estimate in the history of science*" (Though this number in Weinberg (1993) mentioned as $10^{120}$). In the current cosmology models this is known as *the fine tuning problem,* which originally led Weinberg to misleading statement. (For a broader discussion of this approach see reviews of Sahni and Starobinsky 2000; Carroll 2001, 2003; Peebles and Ratra 2003).

### 7.2. The quantum cosmological constant

During the many years a number of papers were published concerning the Einstein cosmological constant problem (see, e.g. McCrea 1987; Sahni and Starobinsky 2000) and this link with the Planck length (Barrow 1988, 1993; Padmanabhan 2003; see, also, Peebles and Ratra 2003). The cosmological constant and Planck parameters have one feature in common: In contrast to quantum-electrodynamics fine structure constant α and the modern theory of strong interaction Fermi's coupling constant $G_F$ they contain the "graviton fine structure constant" $\alpha_{Gr} = Gm_{Gr}^2 / \hbar c$, which is notably small (see, Eq. (8.6) and table 3) and must be brought as a new cosmological conversion factor. In this context notice that, "*the smallness and dimensionless value of* $\Lambda_0 l_{Pl}^2$ *which is still key problem in cosmology*" (Barrow 1988, 1993; Padmanabhan 2003, and references therein) with the high precision might be determined. In the current cosmological models consistent with the classical general relativity the most widely accepted picture is that the cosmological constant not bearing the Planck constant $\hbar$ (see, e.g. Padmanabhan 2003, 2006). Here we recall that a predicted today cosmological constant $\Lambda_0$ can be regarded as a "*Quantum gravity cosmological constant*" because according to Eq. (2.3) and (2.9) this via the constant $\partial$ contains the Planck constant $\hbar$. Carroll (2005) proposed that "*there is not a reliable environmental explanation for the observed value of the cosmological constant*".

At this point, it is interesting to compare the today cosmological constant in terms of the other cosmological parameters estimated above, namely

$$\left.\begin{array}{l} \Lambda_0 \equiv l_0^{-2} \equiv R_U^{-2} \equiv (c^2 / GM_U)^2 \equiv (m_{Gr}c / \hbar)^2 \equiv (m_{Str}c / \hbar)^2 (T_{Pl}^{'} / T_0)^4 \equiv \\ \equiv (T_0^2 / \partial c)^2 \equiv (g_0 / c^2)^2 \equiv (c^4 / 8\pi G) \cdot u_\nu(t_0) \equiv (c^2 / 8\pi G) \cdot \rho_\Lambda(t_0), \end{array}\right\} \tag{7.9}$$

where all parameters hold fixed.

### 7.2.1 Determination of the $\Lambda_0$ from the CMB temperature

As is seen from above set of data, the quantum cosmological constant in terms of Cosmic background temperature have become the subject of experimental study (see, also, Subsec.5.2.3). For example, the correct value $\Lambda_0$ can be precisely determined from experiment as a function of $T_0(CMB) = 2.725(1)K$ (Fixsen and Mather 2002) by using the Eq. (7.9) as

$$\Lambda_0 \equiv (T_0^2 / \partial c)^2 = 1.2915(21) \times 10^{-56} m^{-2}. \tag{7.10}$$

As we have showed in Sec.5, this result is still several orders of magnitude smaller than the current value of the cosmological constant $\Lambda_0 \cong 10^{-52} m^{-2}$ (Peebles and Ratra 2003).

Thus, this anew means dealing with FPC. In this connection, as point out Sahni and Starobinsky (2000) "*cosmology once more becomes a driving force for new insight in physics!*"

## 8. Identifications of the puzzling parameters

It should be notice that the exact value for the today Cosmological constant more than seventy years is the greatest damn mysteries of cosmology and studied in various forms by various authors. First attempts have been made, notably by Eddington and Dirac, to explain this enormous value, but without much success. In units of the Planck length $l_{Pl} \sim 10^{-35} m$ the traditional value $\Lambda_0 l_{Pl}^2$ can be expressed as follows $10^{-119} \div 10^{-123}$ (see e.g. Hawking 1983; McCrea 1987; Barrow 1988, 1993; Padmanabhan 2003, 2006; Peebles and Ratra 2003, and references therein).

If we assume that the Planck length $l_{Pl} = (\hbar G / c^3)^{1/2} = 1.616252(81) \times 10^{-35} m$ (Mohr et al. 2008) then the value $\Lambda_0 l_{Pl}^2$ should be

$$\Lambda_0 l_{Pl}^2 = (l_{Pl}/l_0)^2 = 3.6(4) \times 10^{-126}. \tag{8.1}$$

Padmanabhan (2003, 2006) give considerably large approximately value, i. e.

$$\Lambda_0 l_{Pl}^2 = \Lambda_0 (G\hbar / c^3) \approx 10^{-123} , \tag{8.2}$$

which almost three orders of magnitude more than this in relation (8.1). In this case, a largeness number that is the inverse of value from (8.1), *"which in 1930s was a regarded as a major problem by Eddington and Dirac"* (see e. g. Barrow 1988, 1993, and references cited therein), at present defined by

$$(\Lambda_0 l_{Pl}^2)^{-1} = (l_0 / l_{Pl})^2 = 2.8(3) \times 10^{125}. \tag{8.3}$$

Using the expression for *the Trans-Planck redshift* $z_{TPl} = z_{Pl}^2$ given by (5.36) we identify

$$(l_0 / l_{Pl})^2 = (l_0 / l_{TPl}) \equiv z_{Pl}^2 = z_{TPl}. \tag{8.4}$$

Now, relation (8.1) which, "*today is regarded as a major mystery*" (Barrow, 1993) is found as reciprocal value of the $z_{Pl}^2$

$$(l_{Pl}/l_0)^2 = (l_{TPl}/l_0) \equiv z_{Pl}^{-2} = z_{TPl}^{-1}. \tag{8.5}$$

Then, from a cosmological point of view relations (8.1) – (8.5) may reflect some deep-rooted connection between *Trans-Planck redshift* and other *"mystery"* constants. In the smallness case, Eq. (7.10) combined with Eq. (8.5) in extended version yields

$$\left.\begin{array}{l} z_{TPl}^{-1} \equiv z_{Pl}^{-2} \equiv \Lambda_0 l_{Pl}^2 \equiv (l_{Pl}/l_0)^2 \equiv \rho_\Lambda(t_0)/\rho_\Lambda(t_{Pl}) \equiv \\ \equiv u_\nu(t_0)/u_\nu(t_{Pl}) \equiv \alpha_{Gr} \equiv G m_{Gr}^2/\hbar c \equiv \cdots = 3.59(40) \times 10^{-126}. \end{array}\right\} \quad (8.6)$$

In the largeness case, combination Eq. (8.3) with Eq (8.4) in extended version is

$$\left.\begin{array}{l} z_{TPl} \equiv z_{Pl}^2 \equiv \Lambda_{Pl}/\Lambda_0 \equiv \rho_\Lambda(t_{Pl})/\rho_\Lambda(t_0) \equiv u_\nu(t_{Pl})/u_\nu(t_0) \equiv \\ \equiv (l_0/l_{Pl})^2 \equiv \alpha_{Gr}^{-1} \equiv (m_{Pl}/m_{Gr})^2 \equiv \cdots = 2.79(30) \times 10^{125}. \end{array}\right\} \quad (8.7)$$

Then $M_U/m_{Gr} \equiv N_{Gr} \equiv z_{Pl}^2 = 2.8 \times 10^{125}$ is the number of graviton in the observable universe originated at the Trans-Planck time $t_{TPl}$, so that $\sum_1^N m_{Gr} \equiv M_U \equiv N_{Gr} m_{Gr}$.

The creation of graviton under some circumstances has been investigated by Carr (1980). Their estimate yields $N_{Gr} \sim 10^{120}$ at the Planck time. The dimensionless parameter $\rho_\Lambda(t_0)/\rho_\Lambda(t_{Pl}) = 3.6(4) \times 10^{-126}$, which can be viewed in Eq. (8.6) as $z_{Pl}^{-2}$, elsewhere (Tegmark et al. 2006, Table 1) is accepted as dark energy density corresponding to a $\rho_\Lambda = (1.25 \pm 0.25) \times 10^{-123}$. In other citation (Melchiorri et al. 2007) conditions with $\rho_\Lambda/\rho_r \equiv 10^{-123}$ is dated to the Planck epoch (when the Planck time $t_{Pl} = (\hbar G/c^5)^{1/2} = 5.39 \times 10^{-43} s$, while the appearance of this ratio is appropriate to the time $t_{TPl} = t_{Pl}^2/t_0 \cong 10^{-106} s$ of the Trans-Planck epoch; see e.g. Eq. (5.19c)).

## 9. Current cosmological parameters from the CMB temperature

To conclude, we have summarized our present knowledge about the most important cosmological parameters for the today Universe. In table 3 have been directly determined and gives values of the 15 cosmological parameters. They are the $t_0$, $l_0$, $g_0$, $\rho_\Lambda(t_0)$, $u_\nu(t_0)$, $\Lambda_0$, $M_U$, $m_{Gr}$, $m_e/m_{Gr}$, $m_{Z^0}/m_{Gr}$, $m_{Str}$, $N_{Pl}$, $N_{Gr}$, $N_{Str}$ and $\alpha_{Gr}$. We present results as a function of the two CMB temperatures: $T_0(pred) = 2.766_{25}(160) K$ and $T_0(CMB) = 2.725(1) K$ (Fixsen and Mather 2002).

Throughout this paper we use the values of the $FPC$ $c, \hbar$, $G$, $m_e$, $\tilde{\lambda}_e$, based on the $CODATA$-2006 compilation (Mohr et al. 2008). The parameters of $Z^0$ boson comes from the Particle Data Group ( Tegmark et al. 2006; Yao 2006).

**Table 3**. Current cosmological parameters as a function of the *CMB* temperature $T_0$

| Quantity | Symbol, equation | Value |
|---|---|---|
| background temperature | $T_0 = (\partial c A_{S0})^{-1/2}$ | $2.766(160)K - 2.725(1)K$ |
| age of universe | $t_0 = \partial / T_0^2$ | $[2.85(16) - 2.935(2)] \times 10^{19} s$ |
| horizon size | $l_0 = \partial c / T_0^2$ | $[8.54(49) - 8.799(6)] \times 10^{27} m$ |
| background acceleration | $g_0 = (c / \partial) T_0^2$ | $[1.052(61) - 1.0214\,(4)] \times 10^{-11} m / s^2$ |
| today matter density | $\rho_\Lambda (t_0) = (8\pi)^{-1} (T_0^4 / G \partial^2)$ | $[7.35(43) - 6.920(10)] \times 10^{-31} kg / m^3$ |
| today energy density | $u_\nu (t_0) = \rho_\Lambda (t_0) \cdot c^2$ | $[6.61(40) - 6.219(10)] \times 10^{-14} J / m^3$ |
| today Cosmological constant | $\Lambda_0 = (T_0^2 / \partial c)^2$ | $[1.37_1(16) - 1.2915(21) \times 10^{-56} m^{-2}$ |
| Universe mass | $M_U = \partial c^3 / G T_0^2$ | $[1.15(7) - 1.1849(9)] \times 10^{55} kg$ |
| graviton mass | $m_{Gr} = (\hbar / \partial c^2) T_0^2$ | $[4.11(24) - 3.9976(29)] \times 10^{-71} kg$ |
| electron-graviton mass ratio | $m_e / m_{Gr} = \partial c / T_0^2 \lambdabar_{ce}$ | $[2.21(13) - 2.2787(17)] \times 10^{40}$ |
| $Z^0$ boson-graviton mass ratio | $m_{Z^0} / m_{Gr} = \partial c / T_0^2 \lambdabar_{cZ^0}$ | $[3.95(23) - 4.0663(30)] \times 10^{45}$ |
| String mass | $m_{Str} = m_{Gr} (T_0 / T'_{Pl})^2$ | $[7.78(45) - 7.343(3)] \times 10^{-134} kg$ |
| number of "Planck particles" | $N_{Pl} \equiv z_{Pl} \equiv (T'_{Pl} / T_0)^2$ | $[5.28(31) - 5.444(2)] \times 10^{62}$ |
| number of gravitons | $N_{Gr} \equiv z_{PL}^2 \equiv (T'_{Pl} / T_0)^4$ | $[2.79(30) - 2.964(4)] \times 10^{125}$ |
| number of strings | $N_{Str} \equiv z_{PL}^3 \equiv (T'_{Pl} / T_0)^6$ | $[1.48(26) - 1.614(4)] \times 10^{188}$ |
| fine structure constant of graviton | $\alpha_{Gr} \equiv z_{Pl}^{-2} \equiv (T_0 / T'_{Pl})^4$ | $[3.59(40) - 3.374(5)] \times 10^{-126}$ |

## 10. Concluding comments

As illustrated in figure 1, the synthesis of the Big Bang model with the Freundlich redshift in the very early period of expansion rate of the universe precisely lead to several sharp and discrete phase of transitions with a steps of about $3 \times 10^{62}$ of magnitude. These phase transitions may be separated into two classes: The first class are associated with a three step-like phase transitions between the Over Trans-Planck ($z_{OTPl} \equiv z_{Pl}^3$), Trans-Planck ($z_{TPl} \equiv z_{Pl}^2$) and the Planck $z_{Pl}$ redshifts, where go on concurrently change of entropy (Reif 1965). The kindred redshifts correspond to a transition after the Planck epoch ($Z^0$ boson phase, electromagnetic phase, and etc.), where entropy is growing on at all times. We have calculated these redshift relations, obtaining a good agreement between these steps. It is worth noting that our treatment found no difference from general relativity approach, but there are certainly many open issues and others details which remains for the future considerations.

Finally, we point out that, conception on phase distinction between the cosmological particles from the two "periods" (i.e. at $z_{Pl}^3 \rightarrow z_{Pl}^2$ and $z_{Pl}^2 \rightarrow z_{Pl}$) of the very early Universe it is very impressive, and it remains worthwhile in order better to understand the problem of the nature of the Dark energy. This problem *"is one of the deepest and most exciting puzzles in all of science"* (Turner 2003). What's more, one implication of this consideration is that the present-time relative density of dark energy $\Omega_\Lambda$ (equivalently, the matter density $\Omega_M$) can be well determined from a combination of these red shift parameters. But further investigation in this direction is beyond the scope of this work.

**Acknowledgements**

This work based on part of the programme of the Gravity and Cosmological Research Department of the Shemakha Astrophysical Observatory, National Academy of Sciences of the Azerbaijan Republic. The author is grateful to Professor V. K. Abalakin for careful reading first version of the manuscript and corrections of poor English. He also thanks Dr. M. S. Chubey for help in preparation of the last English version of manuscript.